**Refinement of Bolide Characteristics from Infrasound measurements**

Nayeob Gi[a], Peter Brown[b,c]

[a] Department of Earth Sciences, University of Western Ontario, London, Ontario, Canada N6A 3K7 (ngi@uwo.ca)
[b] Department of Physics and Astronomy, University of Western Ontario, London, Ontario, Canada N6A 3K7 (pbrown@uwo.ca)
[c] Centre for Planetary Science and Exploration, University of Western Ontario, London, Ontario, Canada N6A 5B7

## Abstract

We have detected and performed signal measurements on 78 individual bolide events as recorded at 179 infrasound stations between 2006 and 2015. We compared period-yield relations with AFTAC nuclear period-yield data, finding these to be similar with a slight offset. Scatter in period measurements for individual bolide is found to be caused in part by station noise levels and by attenuation effects with range. No correlation was found between the infrasound signal period and any of bolide height at peak brightness, entry speed or impact angle. We examined in detail three well constrained bolides having energy deposition curves, known trajectories and infrasound detections finding some evidence at shorter ranges that a component of station period scatter is due to varying source heights sampled by different stations. However, for longer-range stations in these three case studies, we were not able to assign unique source heights using raytracing due to large uncertainties in atmospheric conditions. Our results suggest that while source height contributes to the observed variance in infrasound signal periods from a given bolide, range and station noise play a larger role.

## 1. Introduction

The recognition that small (1-20m) near-Earth asteroids (NEAs) may pose a significant impact threat to the Earth (Boslough et al., 2015) has led to a renewed impetus to further our understanding of these small near-Earth objects. Observations of bolides and estimation of their characteristics provide critical clues to physical properties, structure, and the overall NEA population such as the flux of small NEAs impacting the Earth (Borovička et al., 2015). In particular, fragmentation behavior, energy deposition with height and total energy yield when correlated with pre-atmospheric orbits may be used to broaden our understanding of the physical properties, structure, and characteristics of small NEAs both individually and as a population. With the development of the International Monitoring System (IMS) in the late 1990s as part of the Comprehensive Nuclear Test Ban Treaty Organization (CTBTO), infrasound stations have been continuously collecting low frequency sound on a global scale from explosive sources for more than a decade. Among the events regularly detected by the infrasound component of the IMS are bolides (Ens et al., 2012). Infrasound is low frequency sound waves extending from the atmospheric Brunt-Vaisala frequency to the limit of human hearing (0.001-20Hz) (Bedard & Georges, 2000). Infrasound is ideal for remote sensing of bolides as such low frequency acoustic waves do not suffer significant attenuation over long distances, making detection and characterization of bolides at long ranges possible.

The IMS detects these objects as their entry to Earth's atmosphere is accompanied by luminous phenomena (collectively termed a meteor) including heat, ionization and in particular production of a shock. Meteors can produce two distinct types of shock waves. One is a hypersonic (or ballistic) shock wave, which radiates as a cylindrical line source and propagates almost perpendicular to the path of the meteor (Edwards, 2009). A second type of shock is produced when the meteoroid suddenly fragments depositing a large fraction of its total energy over a very short segment of its path. In this case the shock radiates more nearly as a point source (ReVelle, 1974). The detailed theory on the infrasound source of the bolides has been developed by Revelle (1974) and Revelle (1976), however they are most applicable at short ranges (<300km). The existing theory does not take into account intrinsic signal dispersion or effects of the atmospheric turbulence. Therefore, the meteor infrasound signal at large ranges can be modified due to shock wave interactions and conditions in the atmosphere between the meteoroids and the receiver.

In the past, optical observations using photographic, television and video technologies were the dominant techniques used to study bolides (Ceplecha et al., 1998). In the late 1990s, the use of infrasonic technology to register bolides greatly increased due to the establishment of the CTBTO and its implementation of the IMS network. The IMS consists of seismic, radionuclide, hydroacoustic and infrasound stations. The final IMS plan includes 60 global infrasound stations, though only some 45 are installed and operating as of late 2016.

Infrasonic measurements of bolides may provide source location, origin time and an estimate of yield (Edwards, 2009). The yield (or total bolide energy) is of great physical interest as it establishes the scale of the event and forms the most basic property needed to lead to better understanding of both the physical properties of asteroids and their flux at Earth, through ablation modeling. However, past studies using amplitude and particularly period of the infrasonic signal show large interstation variability for common events. Ens et al. (2012) showed that events detected with many stations may have more accurate yields estimated using the dominant period average across all stations, but such averaging is only possible for a limited subset of well observed events. A similar approach has been employed for ground based explosions by the Air Force Technical Applications Centre (AFTAC) which assumes a log normal distribution of periods (Antolik et al., 2014). The root cause of the dispersion in signal periods remains unknown. Possible explanations include:

1. Signals detected at different stations emanating from different positions/heights along the bolide trajectory as suggested by Silber et al. (2009). This could lead to period differences due to different blast radii as a function of energy deposition and/or due to the increased attenuation of higher frequencies for shocks emitted at higher altitudes artificially increasing the apparent signal period.
2. Doppler shifts caused by winds (Ens et al., 2012)
3. Dispersion effects in propagation to different ranges (ReVelle, 1974)
4. Height of burst effects (Herrin et al., 2008; Edwards et al., 2006).
5. Different noise characteristics at each site may alter the apparent period or mask its spectral characteristics (Bowman et al., 2005).
6. Measurement uncertainty and/or broad frequency peaks leading to imprecise dominant period measurements (Golden et al., 2012)

The goal of this study is to examine the infrasonic signals produced by a large sample of bolides, which have known properties as reported on the Web by NASA's Jet Propulsion Laboratory (JPL)[1]. These data are based on U.S. government sensor detections of bolides and report bolide characteristics including location, time, energy, height, speed, and entry angle for a subset of events. Our aim is to explore empirical correlations between measured infrasound parameters at IMS stations (particularly dominant signal period) and bolide secondary characteristics reported on the JPL website. Beyond these empirical explorations, we investigate three test cases in detail to determine if signals emanating from different portions of bolide trails can provide self-consistent explanations for differing signal periods measured at different stations. We wish to test if possibility #1 in the foregoing list is viable explanation for station period scatter. These bolides include the February 15, 2013 Chelyabinsk fireball (Borovička et al., 2013), the September 3, 2004 Antarctica bolide (Klekociuk et al., 2005) and the Park Forest meteorite dropping fireball of March 27, 2003 (Brown et al., 2004). In these cases we have independent estimates of energy deposition and are able to apply raytracing to establish probable source heights. When combined with weak shock modelling (Edwards, 2009), we may then compare predicted dominant periods with observed periods to investigate if different source heights can self-consistently explain the differing station periods. Finally, in the Supplementary Material we provide a database of all our measured infrasound signals extracted from 179 stations representing detections of 78 fireballs (Table S1).

## 2. Theory and background

In previous studies, examinations of the satellite records of bolide detonations were used to estimate the flux of small NEAs (Brown et al., 2002). Their study determined a power-law relationship between the flux and bolide energy, such that a roughly 1m diameter object having a total energy of ~0.1 kiloton TNT equivalent (1kT = $4.184 \times 10^{12}$ J) impacts Earth once every 1-2 weeks. For comparison, a 0.3 kiloton event occurs once every month, a 5 kiloton object strikes annually and one ~50 kiloton object is expected every 10 years (Brown et al., 2002). The IMS is able to detect energies as small as 0.1 kT at multiple stations if wind and geographical location are favorable so we expect ~dozens of bolides to be detected by the IMS annually (Brown et al., 2014).

The speed of a meteoroid impacting Earth is at least 11.2 km/s, though typical impact speeds are closer to 16-18 km/s (Brown et al., 2015). Thus, the geometry of the hypersonic shock cone is well approximated by a cylinder (ReVelle, 1974). The radius of the cylindrical line source, known as the blast radius ($R_o$), is the distance away from the meteoroid trajectory in the atmosphere wherein all of the deposited explosion energy would equal the expansion work required by the weak shock to move the surrounding atmosphere to this radius (Few, 1969). It corresponds approximately to the distance from the trajectory where the shock overpressure equals the ambient atmospheric pressure. Inside the blast radius the atmosphere is strongly shocked leading to non-linear wave propagation (Ens et al., 2012). Using cylindrical line source blast wave theory (Tsikulin, 1970), the blast radius can be calculated as:

$$R_o = \left(\frac{E_o}{P_o}\right)^{\frac{1}{2}} \quad (1)$$

[1] http://neo.jpl.nasa.gov/fireballs/

where $E_0$ is the total energy per unit trail length and $P_0$ is the ambient hydrostatic atmospheric pressure. We apply the ReVelle (1976) weak shock model to calculate the period of wave at ten blast radii by inverting the fundamental frequency of the wave, which is given by:

$$\tau_o = \frac{1}{f_o} = \frac{2.81\, R_o}{C_s} \tag{2}$$

where $R_o$ is the blast radius and $C_s$ is the speed of sound. This implies a power law relation between energy deposition per unit trail length and infrasonic period at a fixed source range, assuming atmospheric pressure is approximately constant.

ReVelle (1974, 1976) presented the first complete theoretical model of meteor infrasound. Extension and observational testing of this early work has included a number of studies, which sought to improve energy estimate accuracy for bolides using both theoretical methods and applications of empirical estimates from man-made ground-level explosions (Edwards et al., 2005, 2006; Ens et al., 2012; Silber et al., 2015). The empirical relations between bolide energy and infrasound properties (notably observed signal period and amplitude) were compared with ground-level explosive sources. The most common practical energy relations used were those produced by the U.S. Air Force Technical Applications Centre (AFTAC) (ReVelle, 1997) which related observed infrasound period to known nuclear explosion yields. The period-yield fits often quoted are:

$$log(E) = 3.34(log\,\tau) - 2.28, \qquad E \leq 200kt \tag{3a}$$
$$log(E) = 4.14(log\,\tau) - 3.31, \qquad E \geq 80kt \tag{3b}$$

where E is energy in kilotons of TNT equivalent and $\tau$ is the infrasound signal period. Fig. 1 shows the original data used to construct the AFTAC period fits. It is apparent that significant scatter in the period per event are present, as similarly shown in Stevens et al. (2002) who presented measured infrasound signals from Soviet atmospheric nuclear tests conducted in 1957 and 1961. They plotted measured period as a function of yield for all data and their result showed substantial scatter in the signal period.

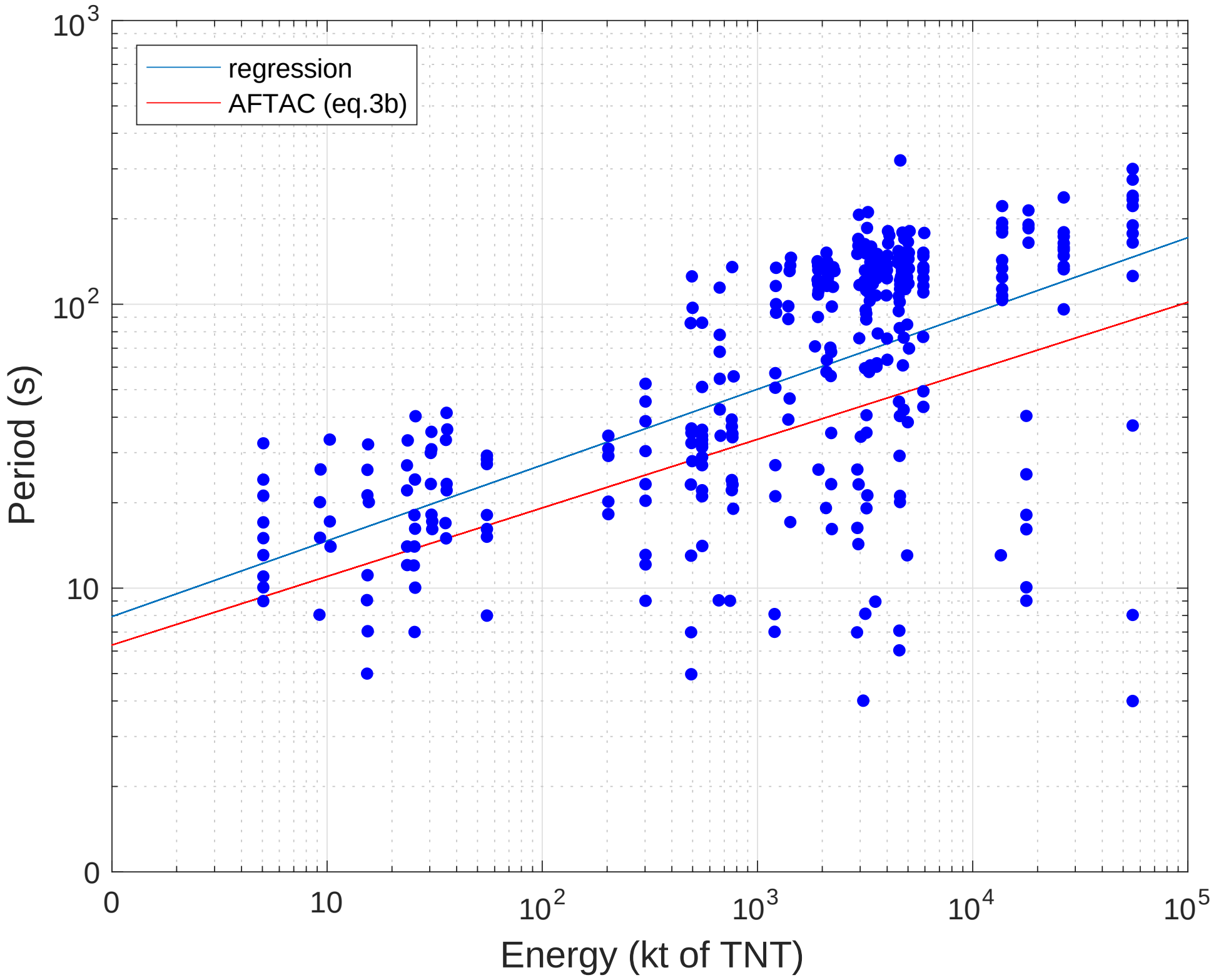

Fig 1: Original AFTAC infrasound data showing yield (x-axis) and individual station periods (y-axis). Note that some of the larger periods are likely Lamb-wave returns (but we do not have sufficient information to determine which of these points are Lamb-wave returns and which are not) and hence were likely not included in the original regression fit (shown in red). A direct regression fit to all data (blue line) produces a similar slope to the AFTAC relation with a vertical shift. Each vertical set of points represents the spread in station periods for one nuclear test.

Historically it has been assumed that the wave period is less modified than the amplitude during propagation. Thus, the period-yield relationship is taken to be more robust (ReVelle, 1997), though the effects of station period scatter have not been systematically investigated. Silber et al. (2011) used global infrasound records associated with the large October 8, 2009 bolide over Indonesia to explore the possibility that different station periods are due to signals emanating from different parts of the bolide trail. They found this to be a plausible explanation for the station period dispersions, but lacked observational ground-truth on the bolide trajectory and energy deposition to make more firm conclusions.

Edwards et al. (2006) showed the theoretically expected change in apparent observed pressure amplitude at the ground scales with source altitude following:

$$\Delta p \propto \left(\frac{P_o}{P}\right)^{\frac{-2}{3}} \tag{4}$$

where $\Delta p$ is the amplitude, $P_o$ is the pressure at ground level, and $P$ is the pressure at the source height. An explosion with yield $E$ at ground level will show a correspondingly larger apparent period as altitude increases as the blast radius (for a fixed energy yield) scales with ambient

pressure – i.e. the blast cavity becomes larger for fixed energy with height. As a result, we expect explosions with constant yields to show larger periods with increasing height, an effect predicted by ReVelle (1976) and observed with small atmospheric explosions (Herrin et al., 2008). The main complication in using signal amplitude is the large corrections needed for the effects of winds, which makes inferring burst heights from amplitudes alone very challenging.

Building on the study of Edwards et al. (2006), Ens et al. (2012) developed empirical relations between bolide total energies as measured by satellites and infrasound signal properties based on a combined study of 71 bolides. Following the same empirical approach as used to generate Eq. (3a/b) they found a power-law relationship between bolide total energy and infrasonic wave period that is linear in log-log space. The best-fit regression to all data was given by

$$log(E) = 3.75(log\,\tau) + 3.50 \tag{5}$$

where $E$ is the satellite-measured bolide kinetic energy in kilotons of TNT and $\tau$ is the infrasonic wave period at maximum amplitude in seconds (Ens et al., 2012). This single station period fit showed considerable scatter. A better fit was found by using multi-station averages for the 30 bolides detected at more than one station. For this multi-station average a fit of the form:

$$log(E) = 3.28(log\,\bar{\tau}) + 3.71 \tag{6}$$

where $\bar{\tau}$ is the multi-station period average was computed.

This is quite close to the AFTAC Eq. (3a), a surprising result as the bolide energy deposition occurs at higher altitudes and hence we would expect larger periods for the same yield. The similar slope (3.28 vs. 3.34) reflects the fact that at the typical large station ranges in our dataset, the finite length of the bolide trail and the effects of atmospheric turbulence lead to the initially cylindrical wave becoming spherical at great distances (ReVelle, 1974). One subtly in comparing the AFTAC and bolide data is that we calibrate the bolide yield from the U.S. government sensors to total radiated energy, which is integrated over the entire path of the fireball. In contrast, the period observed at the ground represents only the energy deposition per unit trail length at some point along the trail, assuming multi-path propagation is not significant. So the correspondence between the AFTAC energies and the bolide total energies may simply be a reflection of the near balancing effects of burst height (which would tend to make the periods appear larger) and energy deposition per unit length (which makes the period appear smaller than if the entire yield occurred at one point).

However, Ens et al. (2012) could not explore the effects of bolide source characteristics on amplitude or period as they were unable to correlate infrasound signals with bolide height, entry angle and speed as these were unknown for their events. We note that while speed is unlikely to greatly affect infrasound characteristics (except indirectly through a correlation with source height) entry angle and the geographical orientation of the trajectory affect infrasonic signal amplitudes, in particular, as recently demonstrated by Pilger et al. (2015). Our work extends the Ens et al. (2012) study by using the NASA JPL fireball data which provides the ground-truth secondary characteristics of bolides, such as height, speed, and entry angle at peak brightness, to explore improvements and limitations in infrasound empirical relationships for bolides related specifically to infrasound period. We note that following Eq. (1) fireballs having the same total energy but different speeds will in general also show different blast radii and hence different periods and amplitudes at the ground.

## 3. Analysis Methodology

3.1 Infrasound signal database construction

We constructed our bolide infrasound signal database cued by the location and timing of ground-truth data from the NASA JPL fireball website. These data, provided by U.S. government sensors, include time, location, height, velocity, total radiated energy, and calculated total impact energy. For each JPL bolide event with all these data, we searched for corresponding signals on infrasound stations of the IMS of the CTBTO. As a guide, we used a probable maximum detection range of:

$$R_{max} = 10^{(\,1.81+0.33\,\log E)} \tag{7}$$

where $R_{max}$ is maximum detection range in km and $E$ is calculated total impact energy in kt of TNT as empirically estimated in Ens et al. (2012). Each potential infrasound waveform from each station was processed using the InfraTool (Fig. 2) component of the analysis package Matseis (Harris and Young, 1997; Young et al., 2002) to isolate the likely bolide infrasound signal.

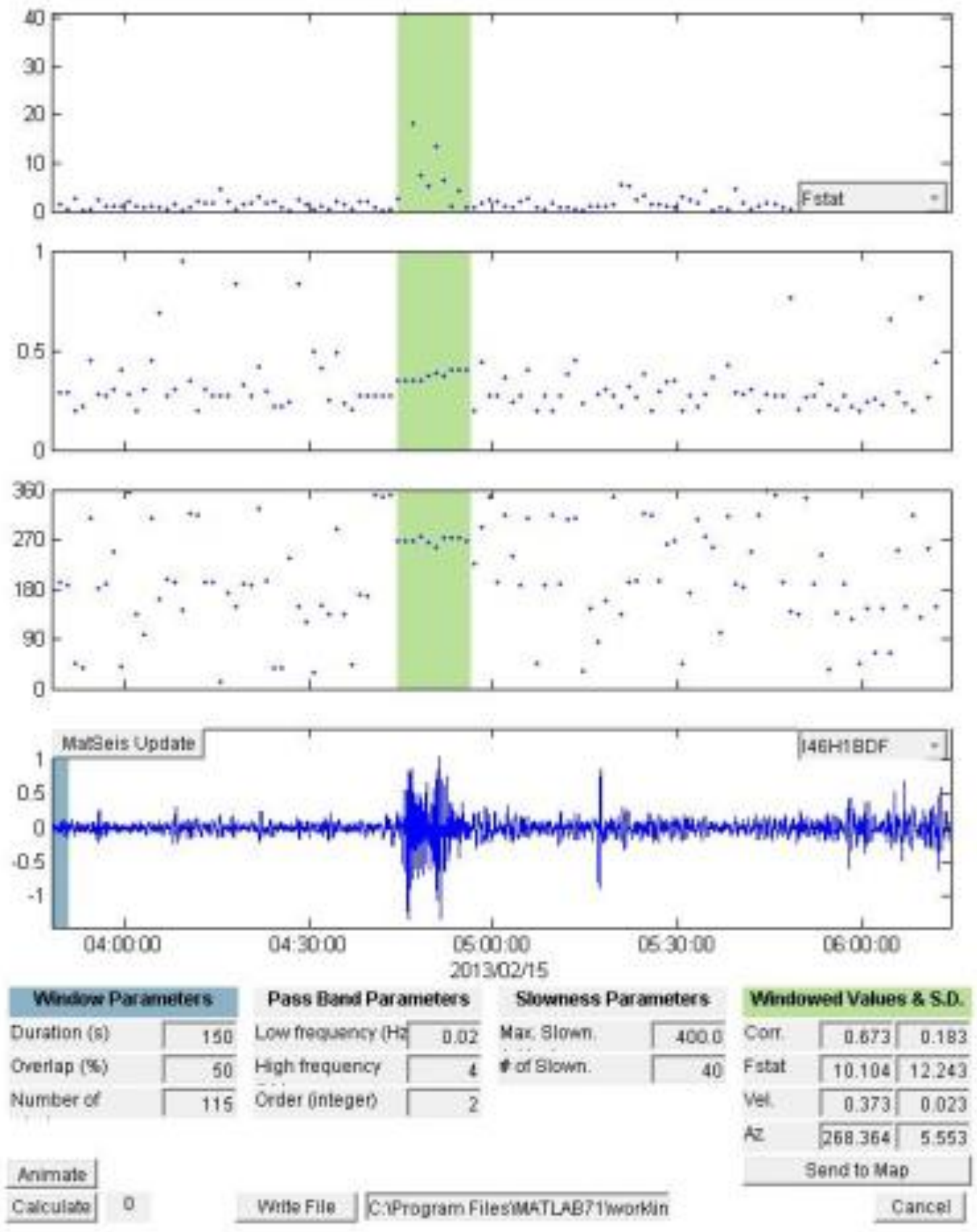

Fig 2: An example screen from the InfraTool analysis package of Matseis (Harris and Young, 1997; Young et al., 2002) showing detection at IS46 for the Chelyabinsk fireball on February 15, 2013. The display window shows (from top to bottom) the F-statistic, trace velocity, back azimuth computed in time windows of 150 seconds duration with 50 % overlap. The final (lowest) graph is raw pressure signal for the first array element. The green region represents the bolide signal where the F-statistic is above the background noise and where the trace velocity and the back azimuth are approximately constant, consistent with what is expected of a coherent infrasound signal.

The InfraTool display windows show the cross-correlation/Fisher F-statistic of waveforms, the trace velocity of the signal, the back azimuth of the waveform, and the waveform filtered by the given frequency bandpass as a function of time. Richard & Timothy (2000) showed that the MB2000, a typical infrasound microphone used at CTBTO stations, has a flat sensor response over 0.01 to 10 Hz. This implies that for signal periods less than 100 seconds (almost all events discussed in this paper including the Chelyabinsk event), the sensor response is flat. The waveform was filtered using a second order Butterworth filter and we have varied the lower and upper cutoff frequency until a maximum signal to noise ratio is achieved. Coherent infrasound signals are first identified by the constant trace velocity and back azimuth values. If there is no signal, the trace velocity and back azimuth will be random because of the continuous fluctuations of pressure produced by winds and other background noise. In cases where the trace velocity and back azimuth change may be indistinct, the cross correlation maximum or the Fisher F-statistic maximum is used to identify the duration of the signal. Appropriate frequency bandpass and window parameters must be chosen in order to maximize the signal-to-noise ratio.

Once an infrasound signal is found, a toolkit termed “inframeasure”, which has specifically been developed for systematic bolide infrasound analysis was used to extract infrasound signal metrics. This process is developed and described in detail in Ens et al. (2012) which built upon the work of Edwards et al. (2006) where the core 7-step process was first employed.

Following this approach, we were able to identify a total of 179 individual infrasound station detections from 78 bolides having complete information on speed, height of peak brightness and entry angle from 2006 to 2015. The detections use InfraTool supplemented with the Progressive Multi-Channel Correlation (PMCC) algorithm as described in Ens et al. (2012). PMCC is an array processing tool that detects coherent signals in frequency and time windows (Cansi, 1995) (Fig. 3). It is sensitive to signals with low signal-to-noise ratio.

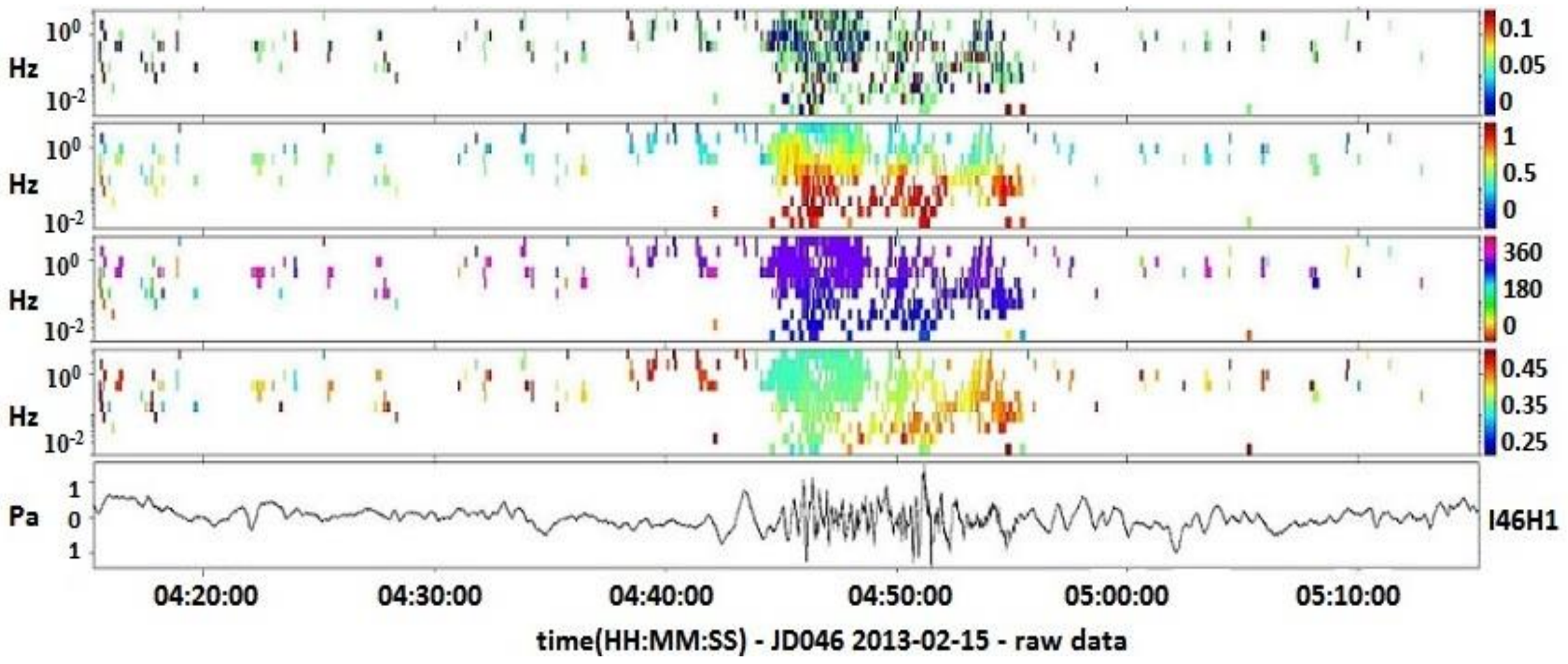

Fig 3: PMCC detection of the infrasound signal at IS46 for the Chelyabinsk fireball on February 15, 2013. The display window shows consistency, correlation, observed back azimuth, trace velocity of the signal, and the raw pressure signal for one of array elements. Details of the PMCC algorithm and its use can be found in Brachet et al. (2010).

## 3.2 Raytracing- GeoAC

We first explored the possibility of identifying source heights of bolide infrasound received at each station for select bolides having sufficient trajectory and lightcurve data through raytracing. Our goal was to find heights which match the observed signal characteristics (arrival times, backazimuth and arrival elevations) and check for self-consistency in terms of predicted periods at the ground using the known bolide lightcurve (and hence energy deposition) coupled with the ReVelle weak shock approach. For this purpose we used the GeoAc raytracing package (Blom, 2014) to extract possible eigenrays, which are individual rays from among a large starting test population of rays at the source which are found to arrive at a given receiver to within some user set distance threshold (in our case 0.1 km).

GeoAc is a numerical package, which models linear acoustic propagation through the atmosphere. In this study, a 3D Range Dependent Global Propagation mode was used which computes ray paths in a three dimensional inhomogeneous atmosphere using spherical coordinates. The atmospheric profile including temperature, pressure, and wind was acquired from the UK meteorological office (UKMO) assimilated data for the altitudes from 0-60km, and for the altitudes above 60km the atmospheric profile was obtained from the Horizontal Wind Model (HWM) (Drob et al., 2015) and US Naval Research Laboratory Mass Spectrometer and Incoherent Scatter Radar (NRLMSIS)-00 model (Picone et al., 2002). The resulting atmospheric splining procedure follows Silber and Brown (2014). We found all eigenrays within all possible inclination ranges and one-degree azimuth windows centred around the great-circle azimuth connecting the bolide source location to the infrasound station (Fig. 4). Because of the large attenuation for thermospherically ducted rays, we are only interested in stratospherically ducted eigenrays. Once we obtained the arrival information for each eigenray, we compared the raytracing results with the observed quantities at the station including travel time, elevation angle, back azimuth and ballistic angle to identify the most probable source height.

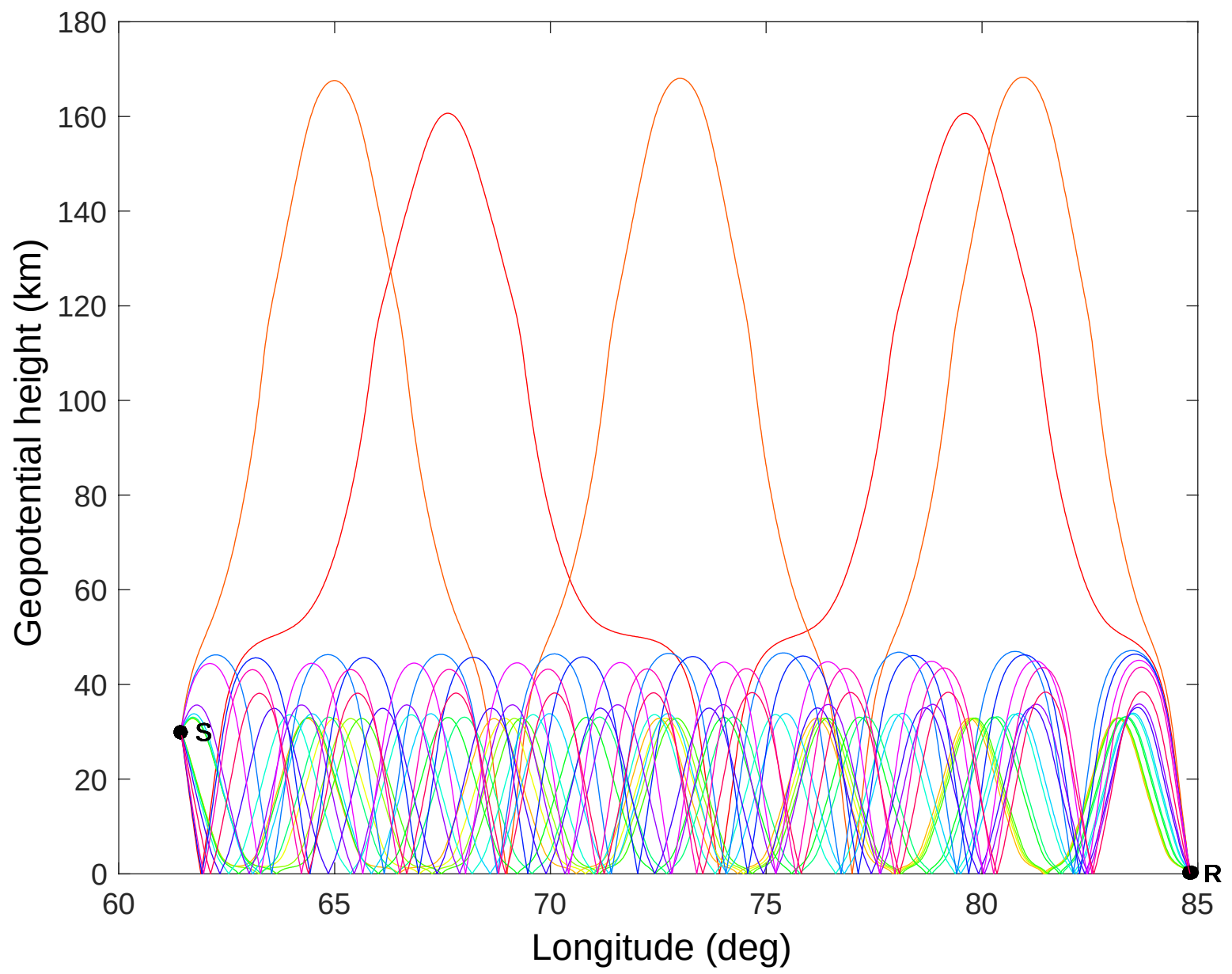

Fig 4: An example of a height – longitude cut along the great-circle path showing a raytracing plot with all eigenrays emitted from a source height of 30km (left hand side of the plot) reaching the station IS46 for the Chelyabinsk event.

### 3.3 ReVelle Weak Shock model

We adapted the ReVelle (1974; 1976) weak shock model to predict what the raytrace estimated source height should produce as a period at the ground in the direction of our measuring stations, given the known energy deposition for several bolides. The weak shock model is an analytical model that requires a set of input parameters characterizing the entry conditions of the meteoroid such as entry angle, trace velocity and blast radius. With the initial conditions, the weak shock model predicts at a given source height what the weak shock period should be at the ground in different directions. In this model, we have the following assumptions:

1. The meteoroid is spherically shaped single body and there is no fragmentation.
2. The trajectory is a straight line therefore gravitational effects are negligible.
3. Only rays that propagate downward and are direct arrivals are considered.
4. Once the transition height is reached (ReVelle, 1976) the wave period remains fixed.

According to the weak shock model, the shock wave reaches its fundamental period after travelling a distance of approximately ten times the blast radius. From this point, the shock wave propagates weakly nonlinearly. According to Towne (1967), the distortion distance (d') is the distance required for a wave to distort by 10% and calculated by

$$d' = \frac{C_s \tau}{34.3\left(\frac{\Delta P}{P_o}\right)} \quad (8)$$

where $C_s$ is the speed of sound, $\tau$ is the signal period, and $\Delta P/P_o$ is the overpressure. Once the shock wave reaches the transition height, the shock is assumed to be in the linear regime ($d' \leq d_a$) where $d_a$ is the remaining distance before a wave reaches the receiver. From this height, the shock propagates linearly and the period remains fixed (Fig. 5). This allows us to compare the predicted infrasound periods with observations. In particular, note that once the linear period is reached, within the assumptions of this model, we have "frozen" the period so detectors at some larger distance will record the same period. In this sense it is possible to examine the predicted period at ranges beyond where direct arrivals only occur, though this is not possible for the amplitude.

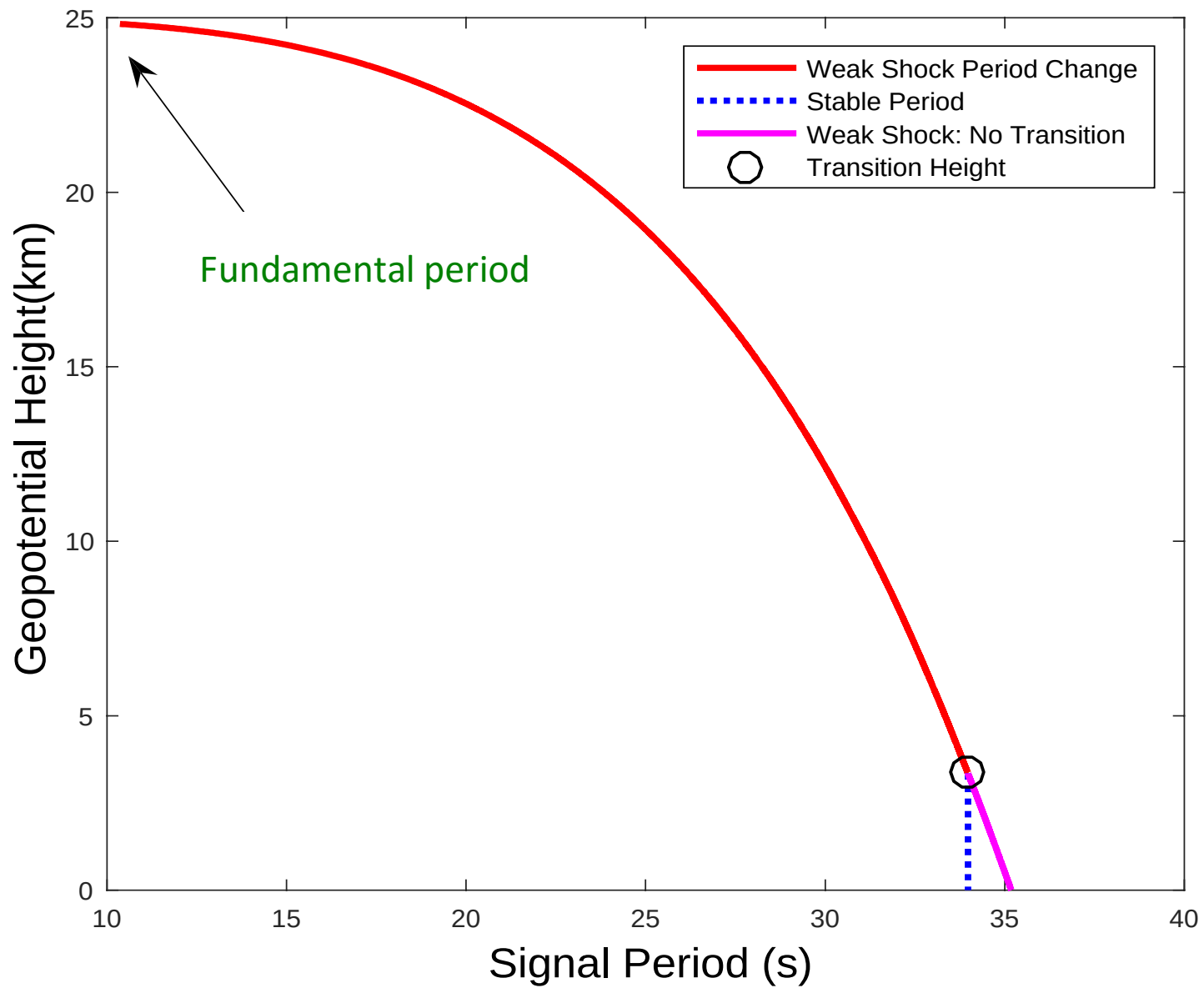

Fig 5: An example of the ReVelle weak shock model result for the February 15, 2013 Chelyabinsk event showing changes in signal period as a function of height in the direction of IS46. The source height was 25km with blast radius 1.08km. In this case, the theory predicts a signal period of 34 seconds.

## 4. Results and Discussion

### 4.1 Period-yield relation

For our study, we have analyzed 78 bolide events as detected from 179 individual infrasound stations from 2006 to 2015 of which 65 events detected from 156 infrasound stations are distinct from Ens et al. (2012). For events that were in common with Ens et al. (2012), we have independently completed inframeasure analysis to compare signal measurements. In a few cases we found some discrepancy between ours and Ens et al. (2012) signal measurements, thus we have re-analysed the bolide infrasound signal with inframeasure several times in order to choose the most appropriate bandpass (which we expect to show the largest SNR) and to remove the background noise. We combined our dataset (Table S1) with the Ens et al. (2012) 50 additional bolide events detected at 143 individual infrasound stations (Table S2) to compare with the period-yield relation from AFTAC (ReVelle, 1997). We also analyzed 37 bolide events that were detected at multiple-stations by taking the average of all signal periods detected at individual station and using one mean period per event. The regression to the combined dataset for all individual detections and multi-station detections were found to be respectively:

$$log(E) = 3.68\,(log\,\tau) - 1.99 \qquad (9a)$$

$$log(E) = 3.84\,(log\,\bar{\tau}) - 2.21 \qquad (9b)$$

where $E$ is the source energy in kt of TNT, $\tau$ is the observed period at maximum amplitude in seconds, and $\bar{\tau}$ is the averaged signal period for a given event detected at different stations. The regression for the combined dataset was found to be very close to the AFTAC period-yield relation as shown in Fig. 6a/b. For any single bolide event, different stations show a large spread in observed periods, similar to the spread in the original AFTAC nuclear explosion data. In principal, we expect a one to one relationship between the period and energy. However, bolides produce cylindrical line source shock along their entire trail (ReVelle, 1976), thus the period measured at each station can be different simply because returns correspond to the size of the cylindrical blast cavity at that particular segment of the trail having an acoustical path to each station. The actual length of segment of the trail that contributes to the signal is unclear since the length depends on non-linear bending near trail (Brown et al., 2007). So one possibility for this large variation is that signals are coming from different part of the bolide trail. Moreover we expect (in the absence of height effects) the line segment sampled at any one station to have an energy deposition only a small fraction of the total bolide energy.

a.

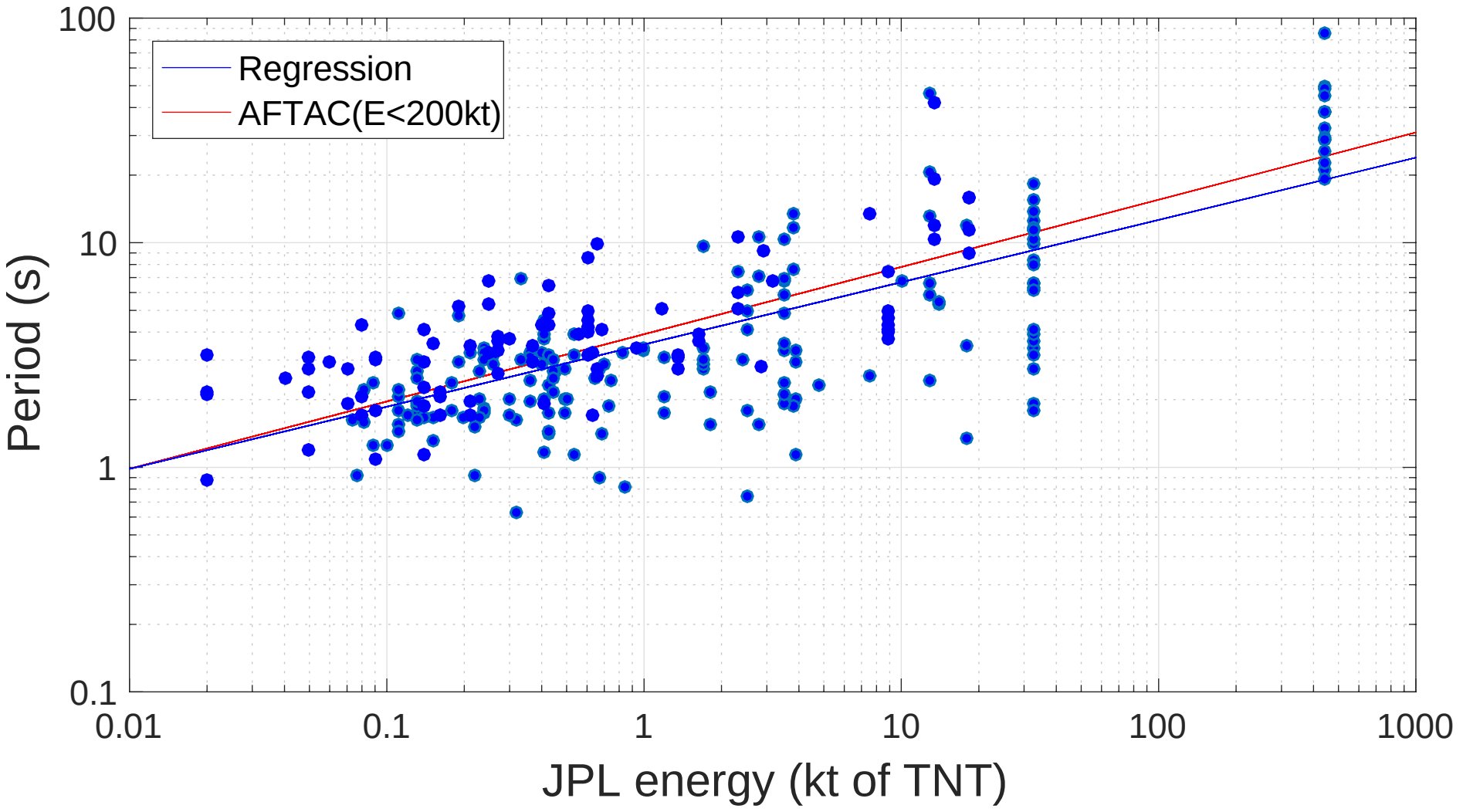

b.

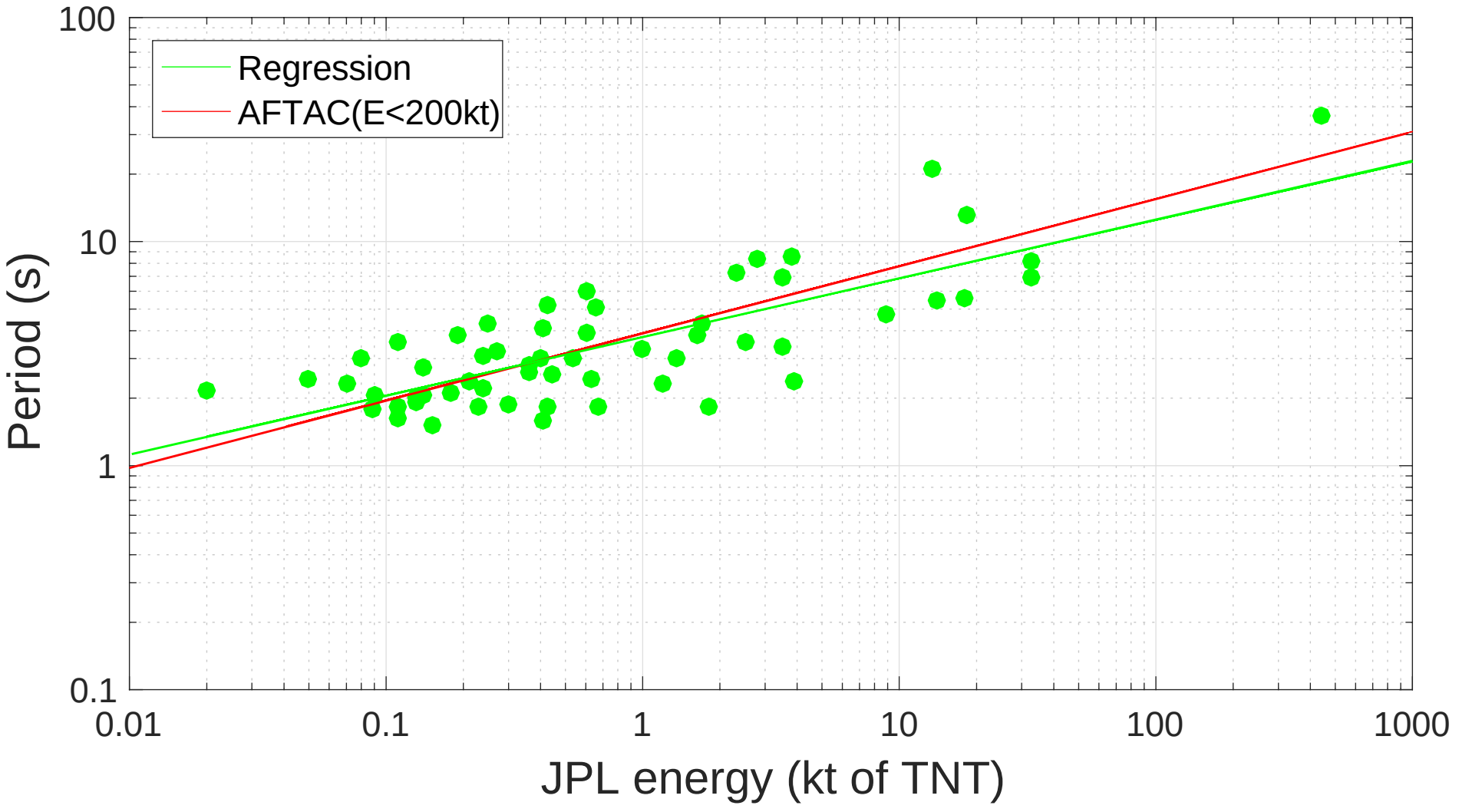

Fig 6: The infrasound signal period at the maximum amplitude as a function of bolide energy as given on the JPL webpage. (a) Each point represents the observed signal period at one particular station. (b) Each point represents the averaged signal period observed for a given event at different stations.

Before exploring this source height effect in detail for specific cases, we first look at our dataset as a whole. Fig. 7 shows the signal period at maximum amplitude as a function of JPL energy for multi-station events color coded by range (km) from the bolide location to different infrasound stations. Longer range stations show higher signal period; In Fig. 7 at fixed energy, there is a weak trend of larger periods at longer ranges. This trend is clearer especially for events > 1kt of TNT. We expect the frequency dependent attenuation to remove higher frequencies with increasing propagation distance (e.g. Norris et al., 2010). This should tend to increase the signal period as the wave propagates further from the source, basically the effect we see in Fig 7.

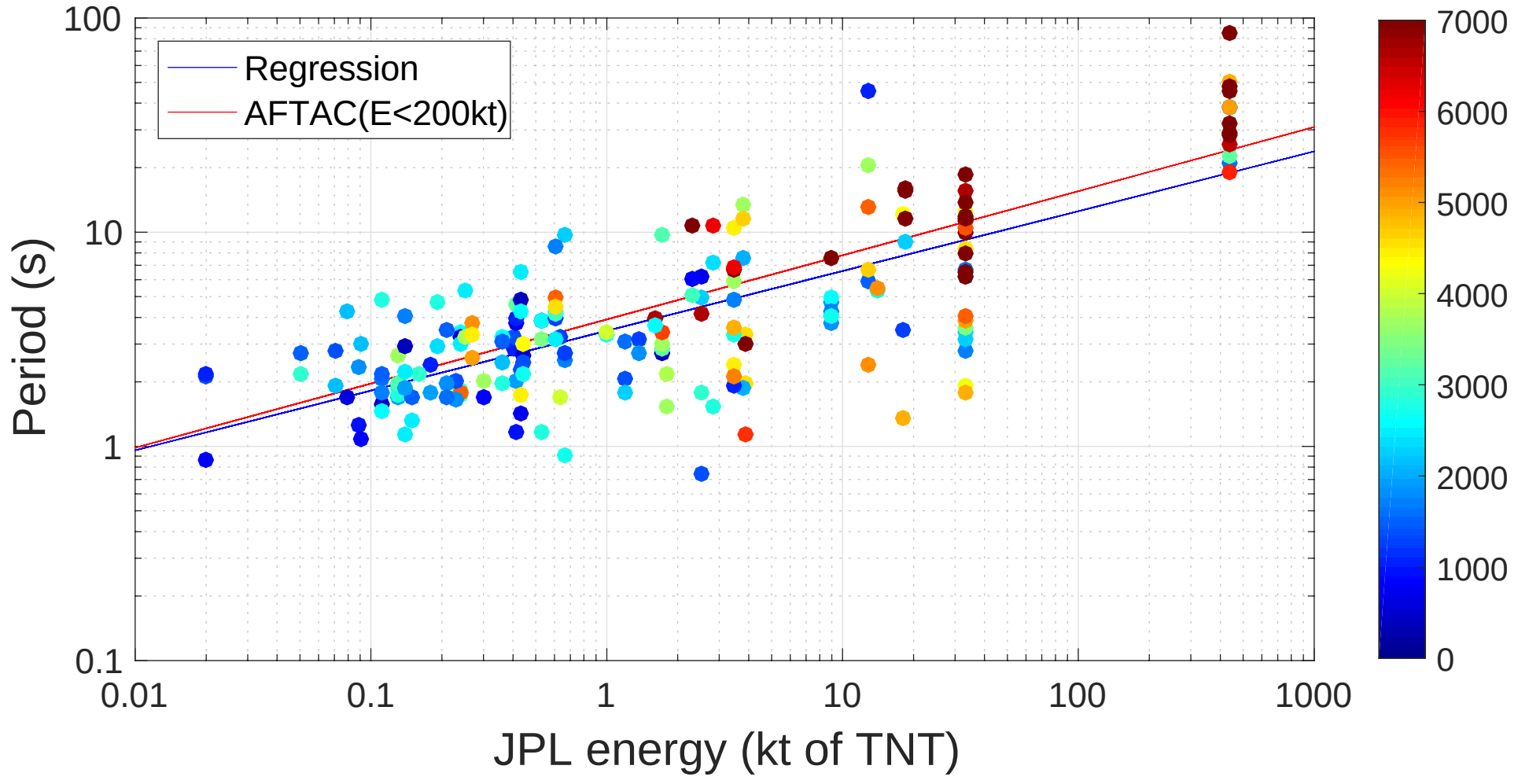

Fig 7: The signal period observed at different stations for multi-station bolide events as a function of JPL energy. Color represents the range (km) from the bolide location to the infrasound station.

Fig. 8a/b/c shows the averaged signal period as a function of JPL energy with color coding for different bolide entry speeds (km/s), bolide heights (km) at peak brightness, and bolide entry angle (degrees). The speed itself is a variable that we do not expect to make a large difference to infrasound period; as expected we do not see any strong correlation. For a given small range in energy, all other things being equal, we would expect to see a vertical gradient in the points whereby the lowest heights show the smallest periods, if the infrasound signals at all stations were predominantly being emitted at the height of peak brightness. The height at peak brightness shows no such strong correlation, though the number statistics in this multi-station average are small (only 37 bolides). This implies that the location along the trail where peak brightness occurs is likely not where the infrasonic periods originate, that each stations sees a different part of the trail and/or the light curve for each event is quite different. No correlation was found between the infrasound signal period and the entry angle.

The simplest interpretation, that infrasound does not dominantly come from where the fireball is brightest, would imply either that individual station-bolide geometries dominant the process or that multiple fragmentation points at different locations in the stratospheric waveguide channel may play a larger role in funneling acoustic energy to a given station. We do not have sufficient information for most of these events to explore this further other than to conclude that burst height does not dominate the observed periods.

a.

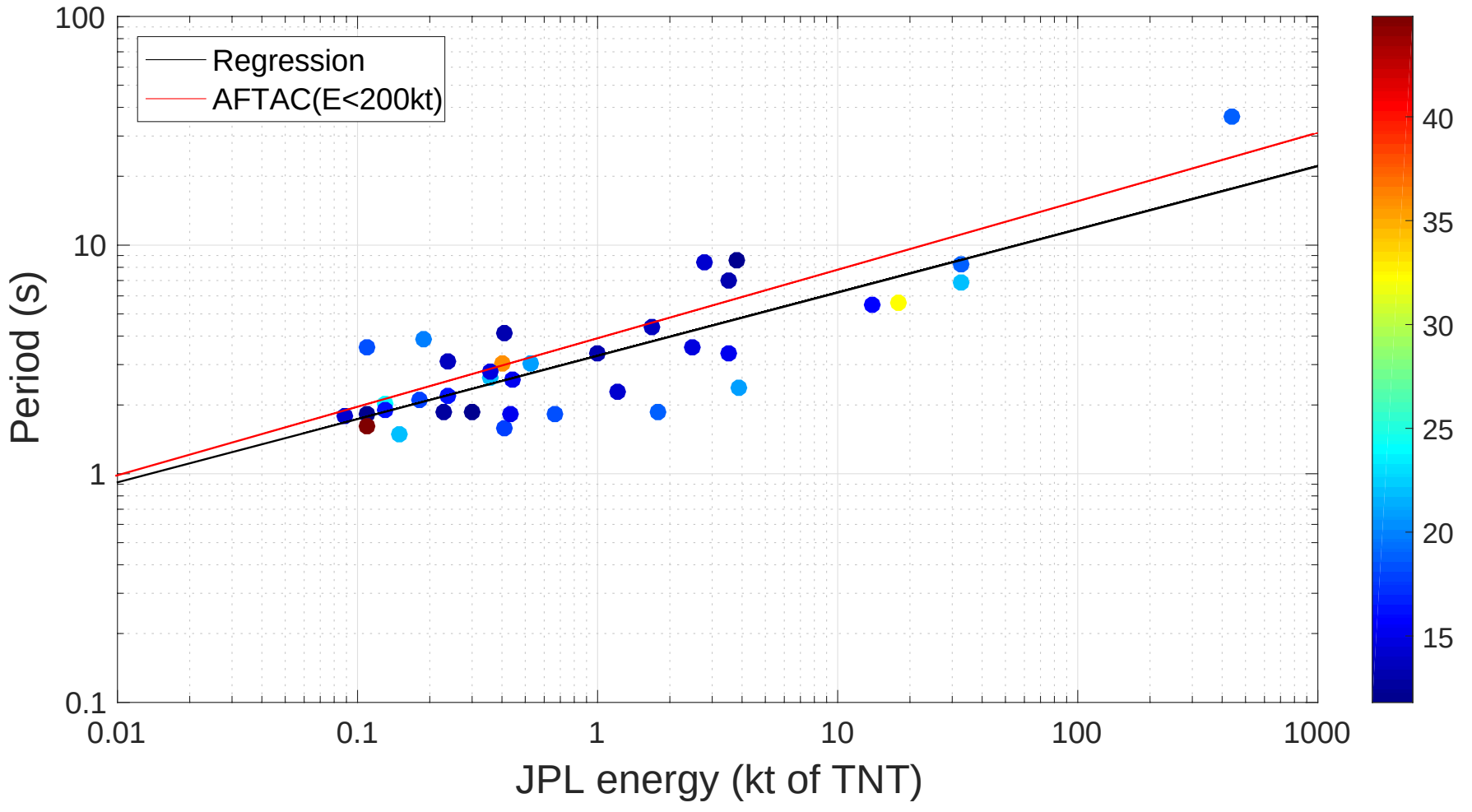

b.

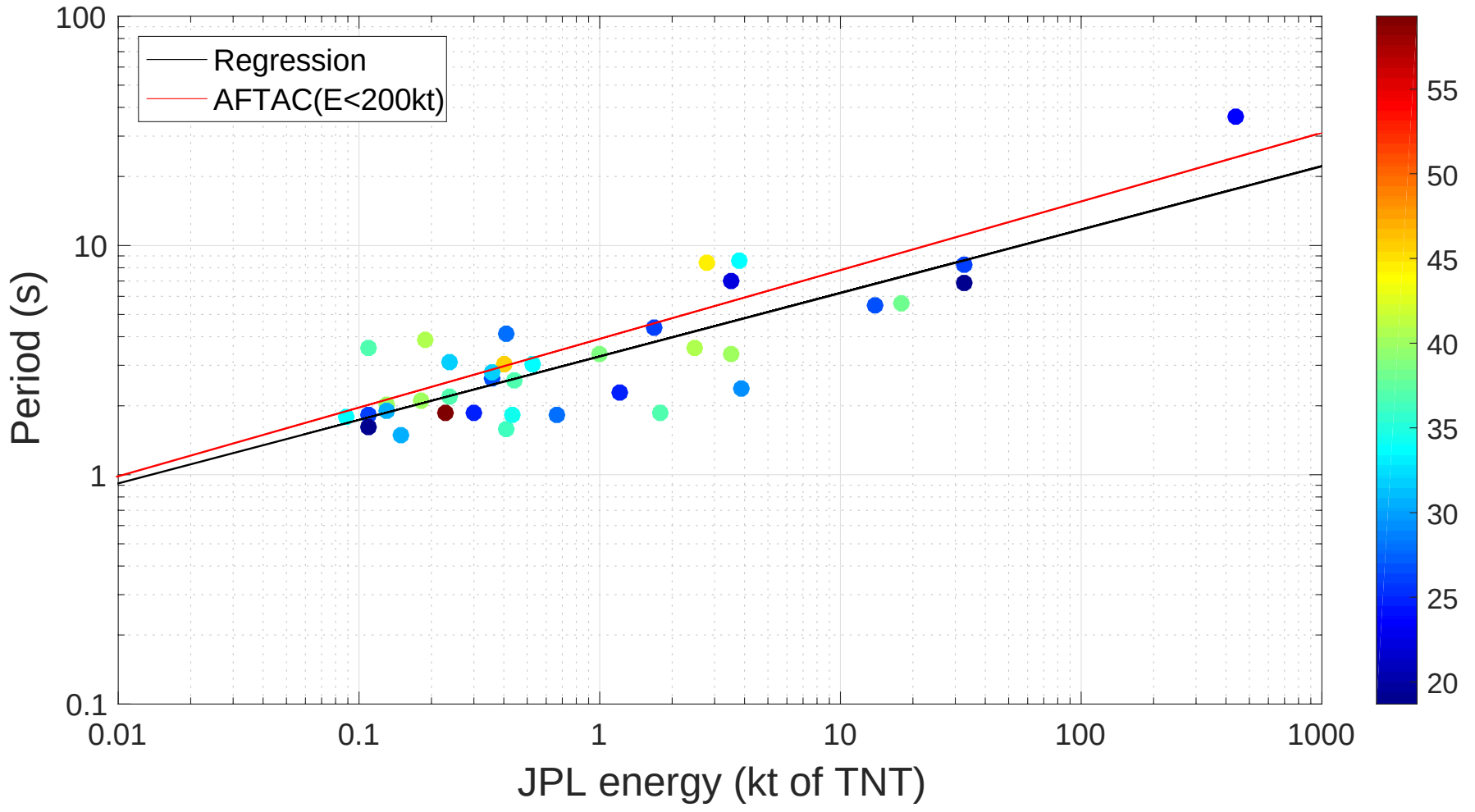

c.

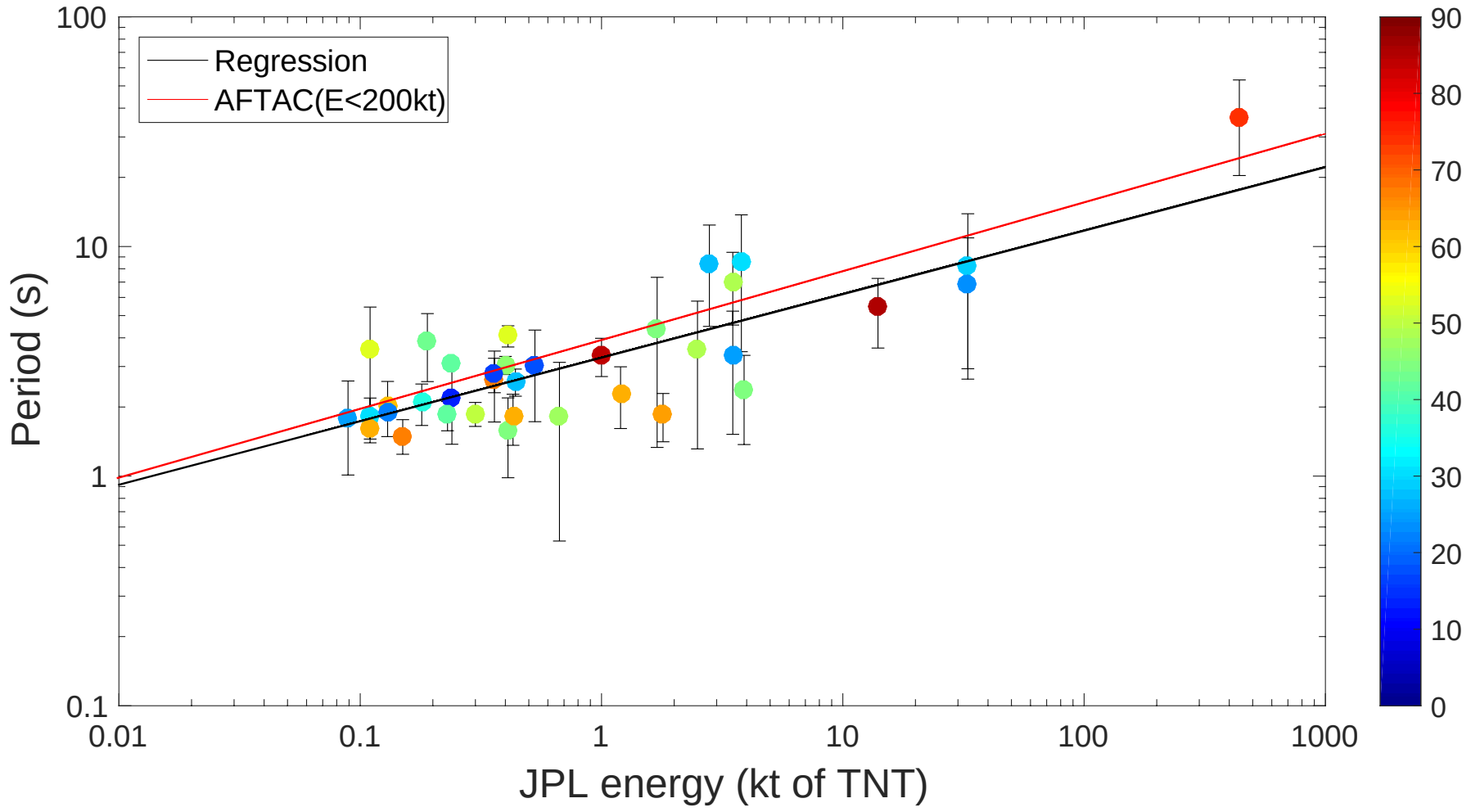

Fig 8: The averaged signal period observed at different stations for multi-station events as a function of JPL energy. (a) Color represents the speed (km/s) at peak brightness. (b) Color represents the height (km) at the peak brightness. (c) Color represents the entry angle (degrees) and the error bars the standard deviation among signal periods measured at the various stations per events.

Pilger et al. (2015) found that station noise levels were the dominant factor in infrasound detection of the Chelyabinsk fireball. Motivated by that study, we also examined the possibility that the peak-to-peak amplitude signal to noise ratio (SNR) plays a significant role in the spread in signal period. As shown in Fig. 9, high SNR points are indeed more clustered along both the AFTAC period-yield relation and our regression to the bolide signal period directly weighed by signal to noise ratio. This trend is clearer as we increase the SNR cut-off value. We have measured the spread of the fit around the AFTAC period-yield relation by calculating the sum of squared residuals (SSR) and the value decreases as we increased the SNR cut-off value. This suggests that SNR is a contributing factor in the dispersion of periods. However, we see that even at high SNR individual events detected at different stations show some spread (though much less than is the case for low SNR station detections). This suggests a more explicit examination of the role of source heights and period is required for cases where sufficient information is available to allow such comparisons.

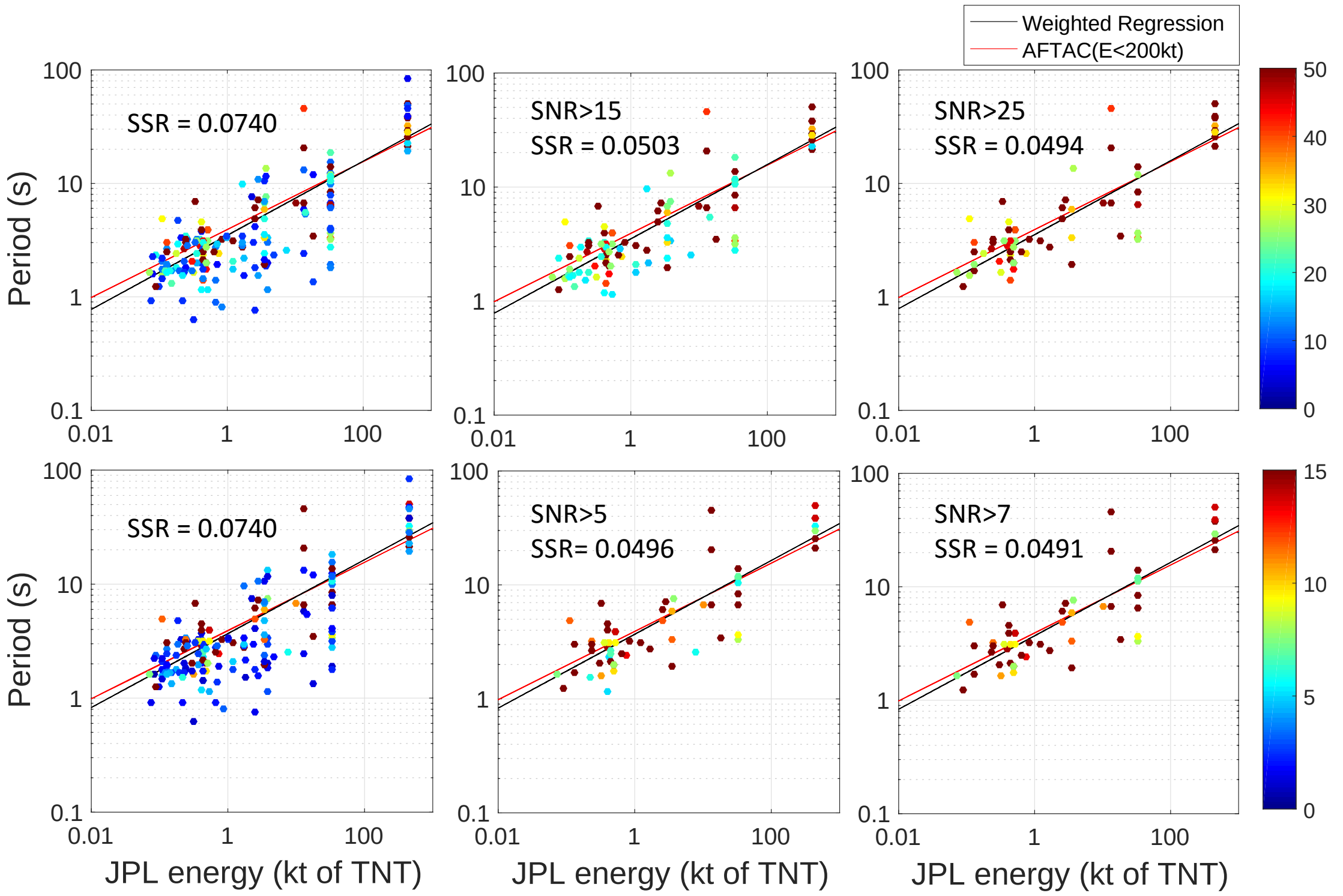

Fig 9: The signal period as a function of JPL energy for all station-bolide detections. The color coding of the upper row plots are the peak-to-peak amplitude signal to noise ratio (SNR) with two different cut off values, SNR>15 and SNR>25. The color of the bottom row plots are the integrated total bolide infrasound waveform energy signal to noise ratio (SNR) with two different cut off values, SNR>5 and SNR>7.

## 4.2 Bolide Infrasound Source Height Estimation: Case studies

To investigate in more detail the possibility that differing source heights may be responsible for the range of periods we investigate three bolides with well-documented trajectories and energy deposition curves.

### 4.2.1 The Chelyabinsk fireball – Feb 15, 2013

Our first case study event was the Chelyabinsk fireball, which occurred on February 15, 2013 at 3:20:33UT over Chelyabinsk, Russia (Borovička et al., 2013; Brown et al., 2013; Popova et al., 2013). This unusually energetic event was detected at over twenty global infrasound stations. We focus on the nearest stations, knowing that our raytrace modelling becomes more uncertain with range.

We were not able to model any stratospherically ducted eigenrays reaching IS31, the closest infrasound station, potentially due to atmospheric uncertainties or counter-wind returns, which are notoriously difficult to model (de Groot-Hedlin et al., 2009). Thus we applied the raytracing-source height technique to the second closest station, IS46. We were able to find eigenrays to this station and established source height by comparing the raytracing model predictions to the observed parameters of signal travel time, elevation angle, back azimuth, and ballistic angle. The results are shown in Fig. 10. Using the travel time and the elevation angle, we isolated the most probable source height as 30km. In this particular case, we could not distinguish source height based on the back azimuth, which shows a large deviation compared to other events (e.g. Ens et al. (2012) data showed azimuth agreement within 10 degrees). This significant deviation is possibly due to the turbulence or because the shock has relatively high amplitude thus, the wave front is more distorted and it is no longer a plane wave.

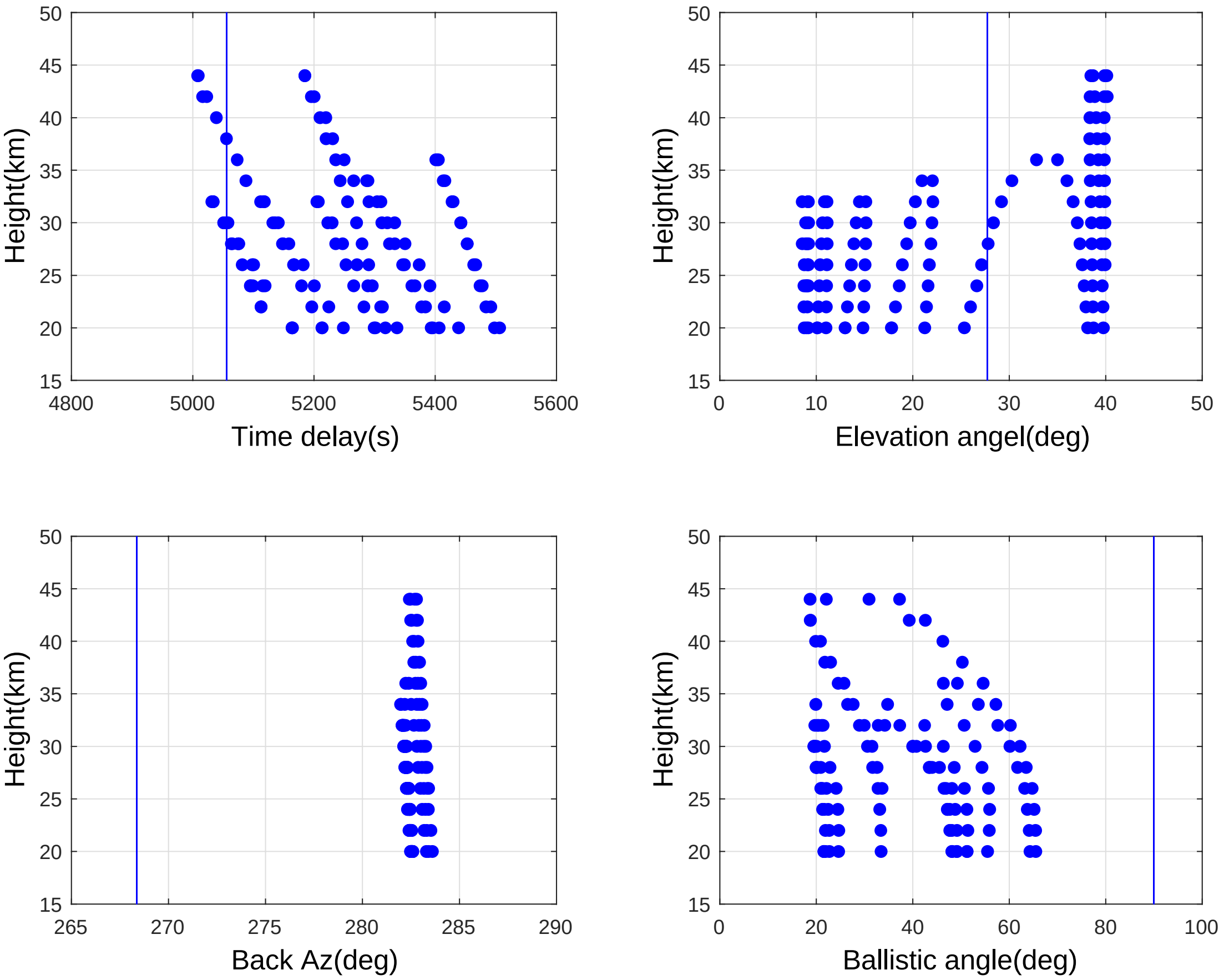

Fig 10: A composite plot showing the travel time (top left), elevation angle at arrival (top right), back azimuth at arrival (lower left) and ballistic angle (lower right) for the February 15, 2013, Chelyabinsk fireball event. The points represent each modelled arrivals from the raytracing at one km height intervals while the vertical solid line corresponds to the observed quantity from inframeasure analysis. The lower right plot shows the take-off angle for the ray from the source where the ballistic angle is the solid line at 90°, expected from a cylindrical line source.

Having established ~30 km as the most probable source height using the raytracing method, we applied the ReVelle weak shock model. Using the energy deposition curve (Fig. 11), we calculated the blast radius as shown in Fig. 12. Chelyabinsk produced blast radii up to about 9.5km at the expected source height 30km, much larger than any other bolide with instrumental measurements. Unfortunately, this blast radius is comparable to or larger than the atmospheric scale height at this altitude, which invalidates one of the assumptions in the use of the ReVelle model. Fig. 13 shows the result from weak shock modeling with source height from 20 to 45km as a function of signal period. In fact we see the greatest disagreement between the simulated signal period and the observed signal period at 30km height. However, this disagreement is not significant because we are violating the assumption that the blast radius is much smaller than the scale height of the atmosphere implicit in the weak shock model. Therefore, we concluded that the period prediction for the ReVelle weak shock model is not valid for the Chelyabinsk event.

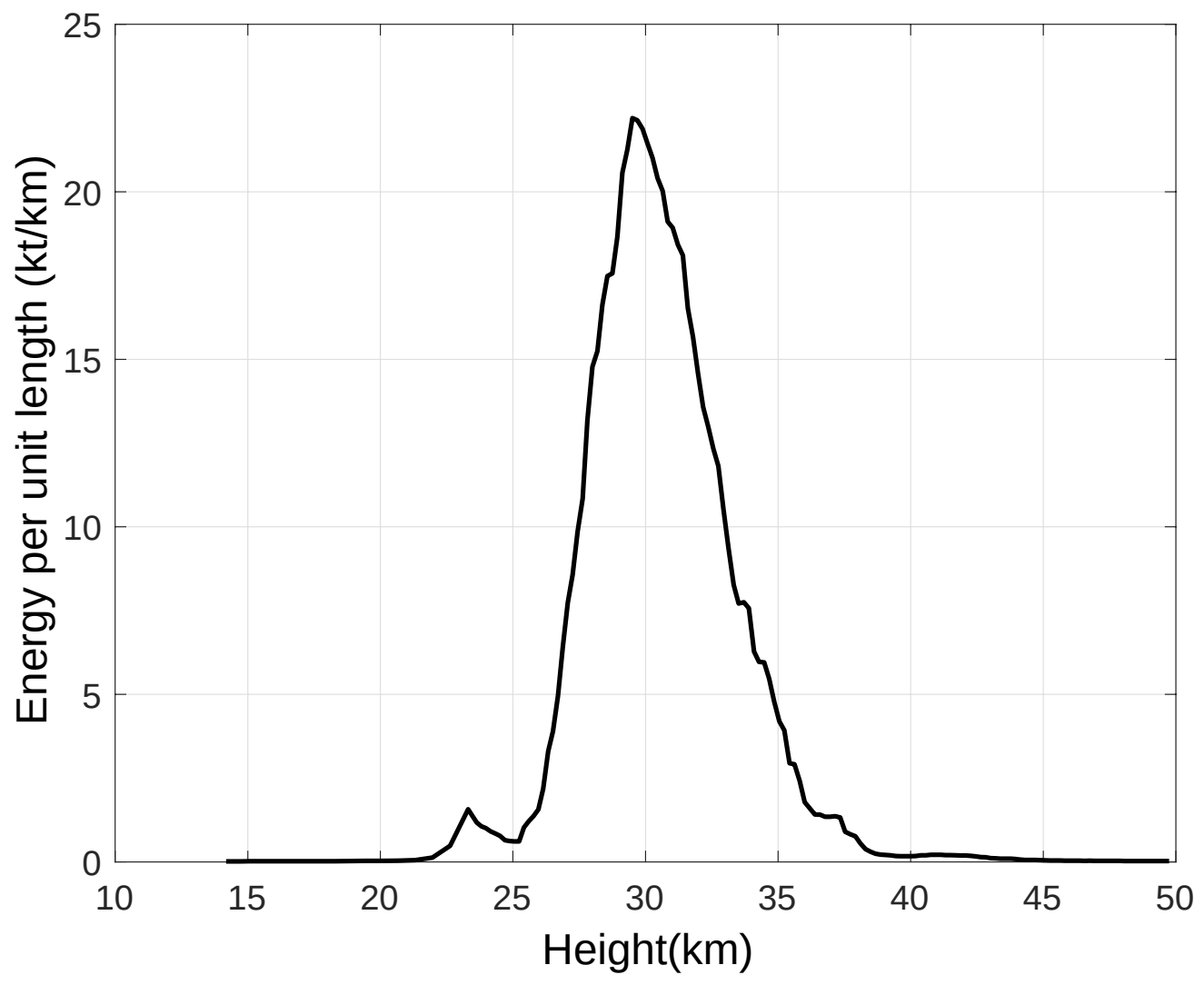

Fig 11: Energy deposition curve for the February 15, 2013 Chelyabinsk fireball taken from Brown et al. (2013).

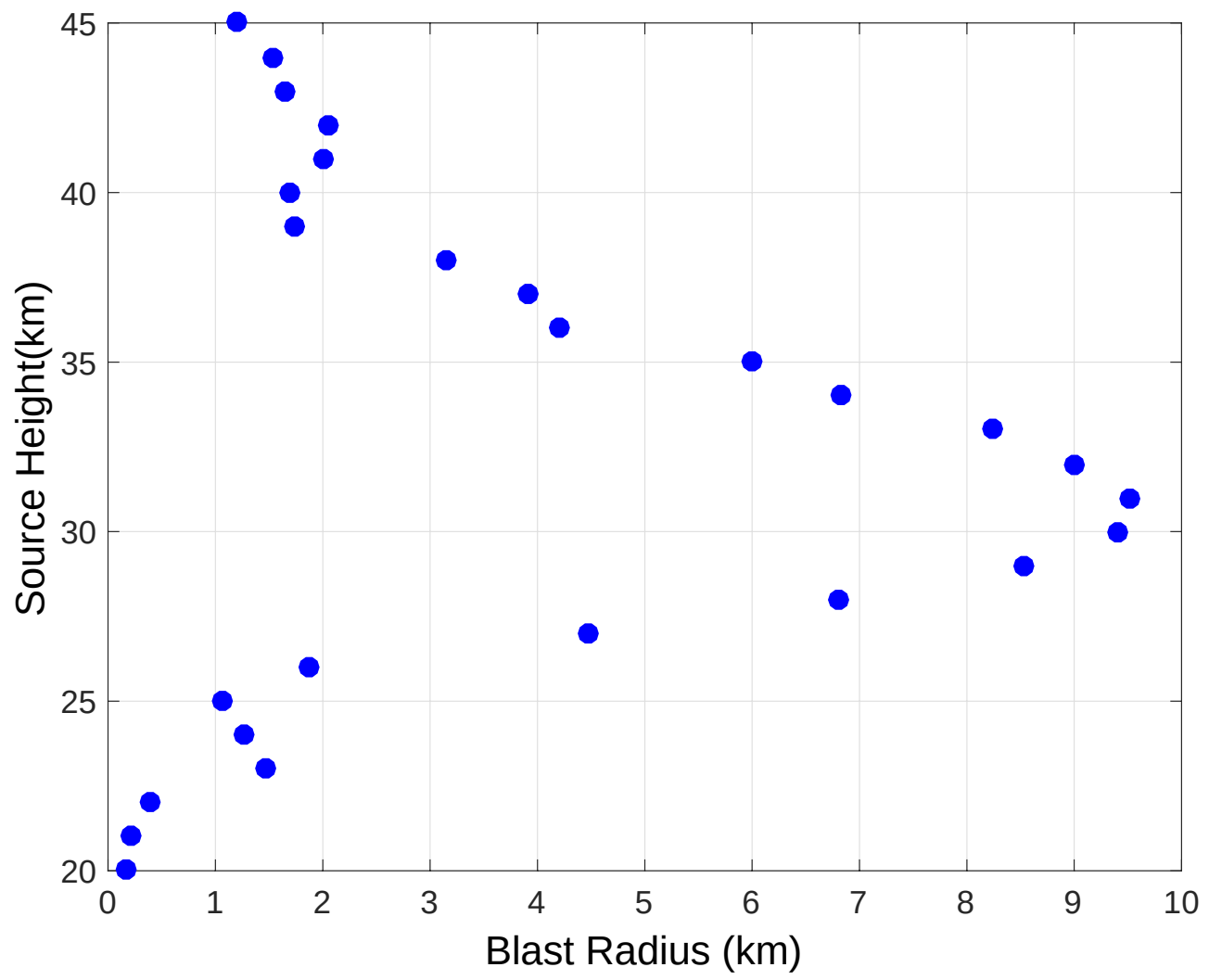

Fig 12: Equivalent blast radius plot for the February 15, 2013 Chelyabinsk fireball based on Fig. 10.

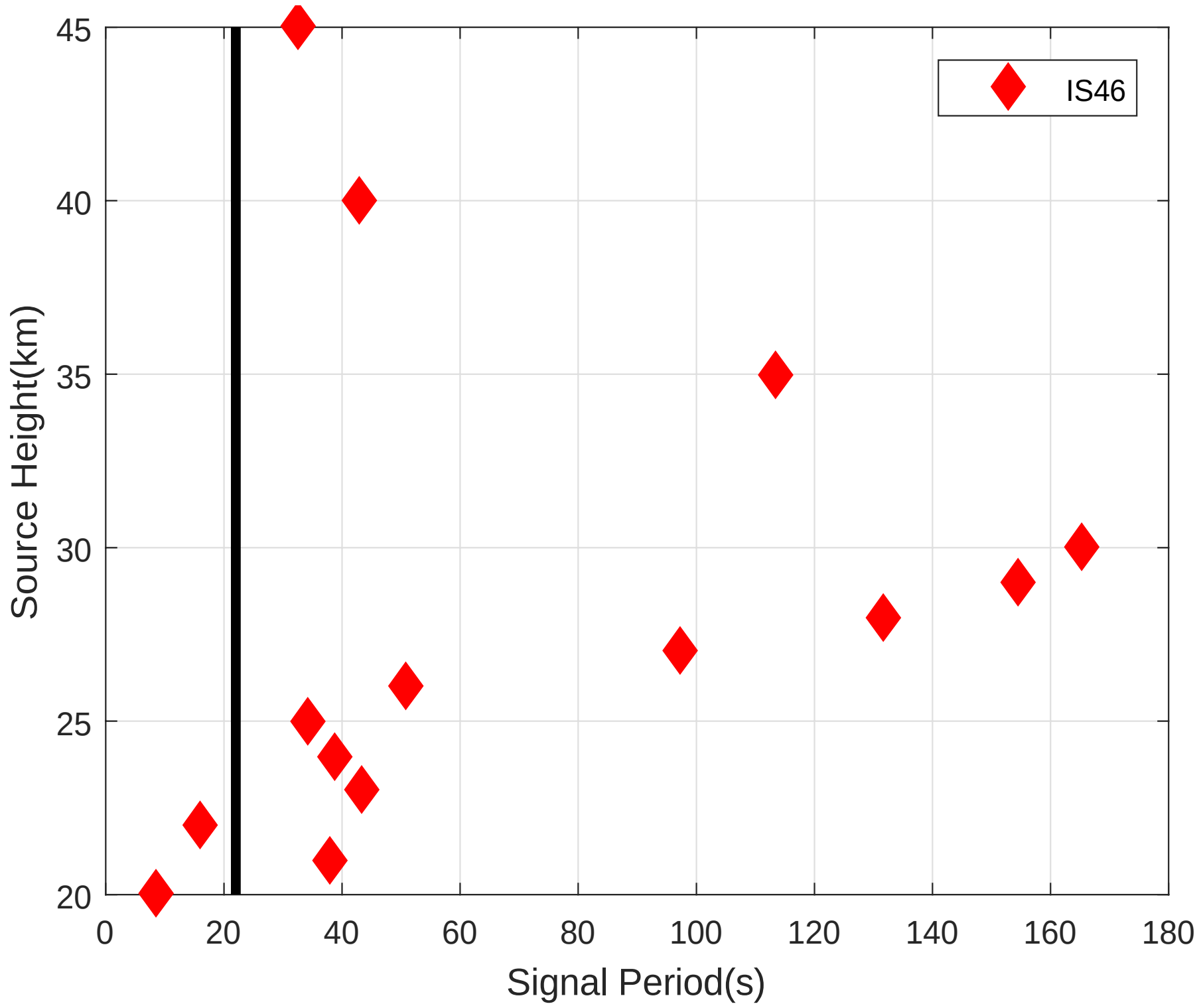

Fig 13: The signal period plot for the Chelyabinsk fireball of February 15, 2013. The diamonds represent simulated signal period from weak shock model and the solid line corresponds to the observed infrasonic signal period (see Table S2 in Supplementary Material).

### 4.2.2 Antarctica fireball – September 3, 2004

Our next case study was a fireball occurring near Antarctica on September 3, 2004 at 12:07:22UT (67.64°S 18.83°E). This event was detected at three infrasound stations, IS27, IS55, and IS35 (Klekociuk et al., 2005). The details of infrasound measurements at each station can be found in the Table S2 of the Supplementary Material. From the U.S. government sensor observations, the total radiated energy is $7.26 \times 10^{12}$ J suggesting total impact energy of 13kT, using the Brown et al. (2002) optical energy-total calibrated energy relation. The light curve from Klekociuk et al. (2005) is shown in Fig. 14 and from the equivalent energy deposition curve, we generated blast radii as a function of height for the event as shown in Fig. 15.

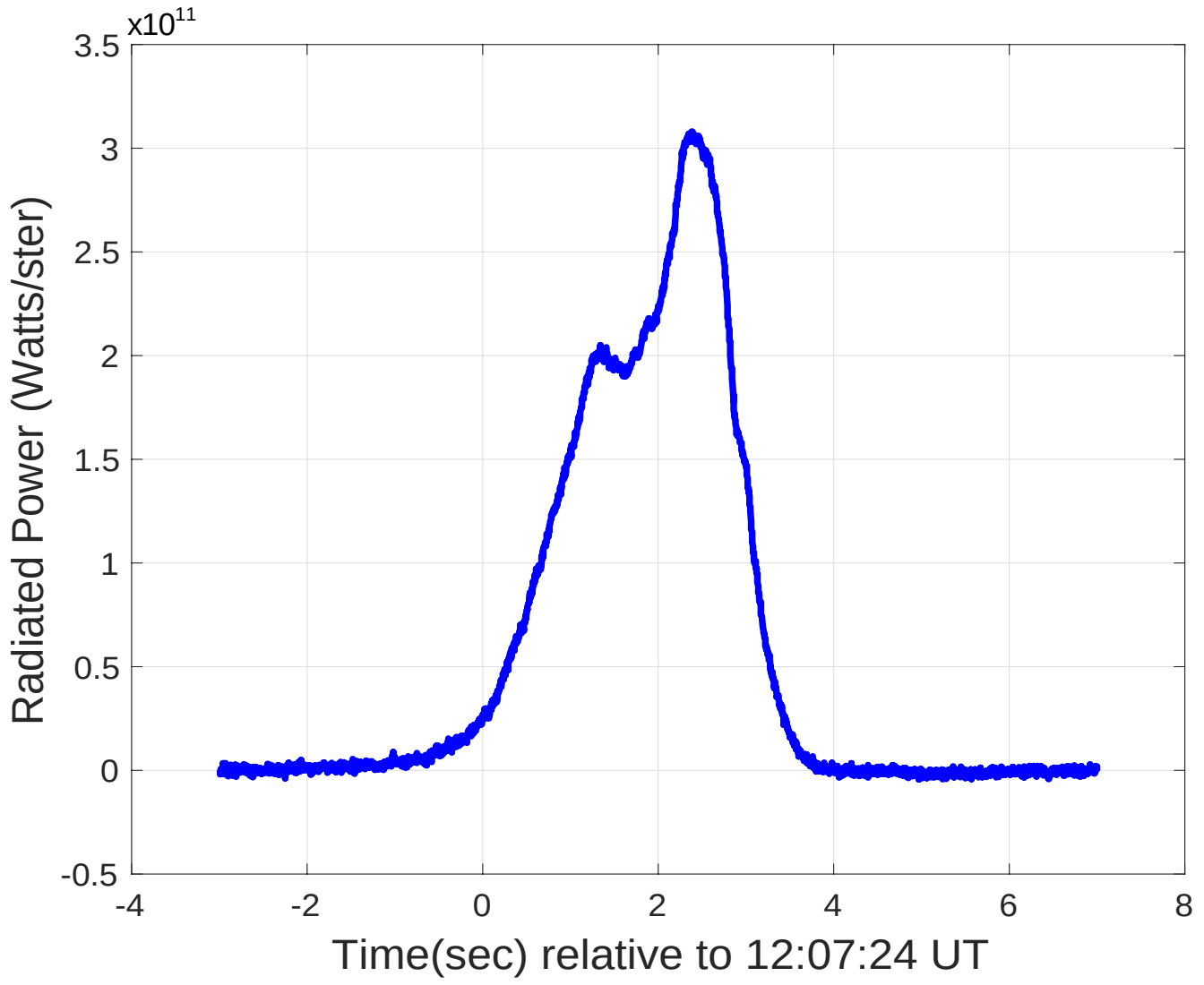

Fig 14: Light curve for the September 3, 2004 Antarctica fireball

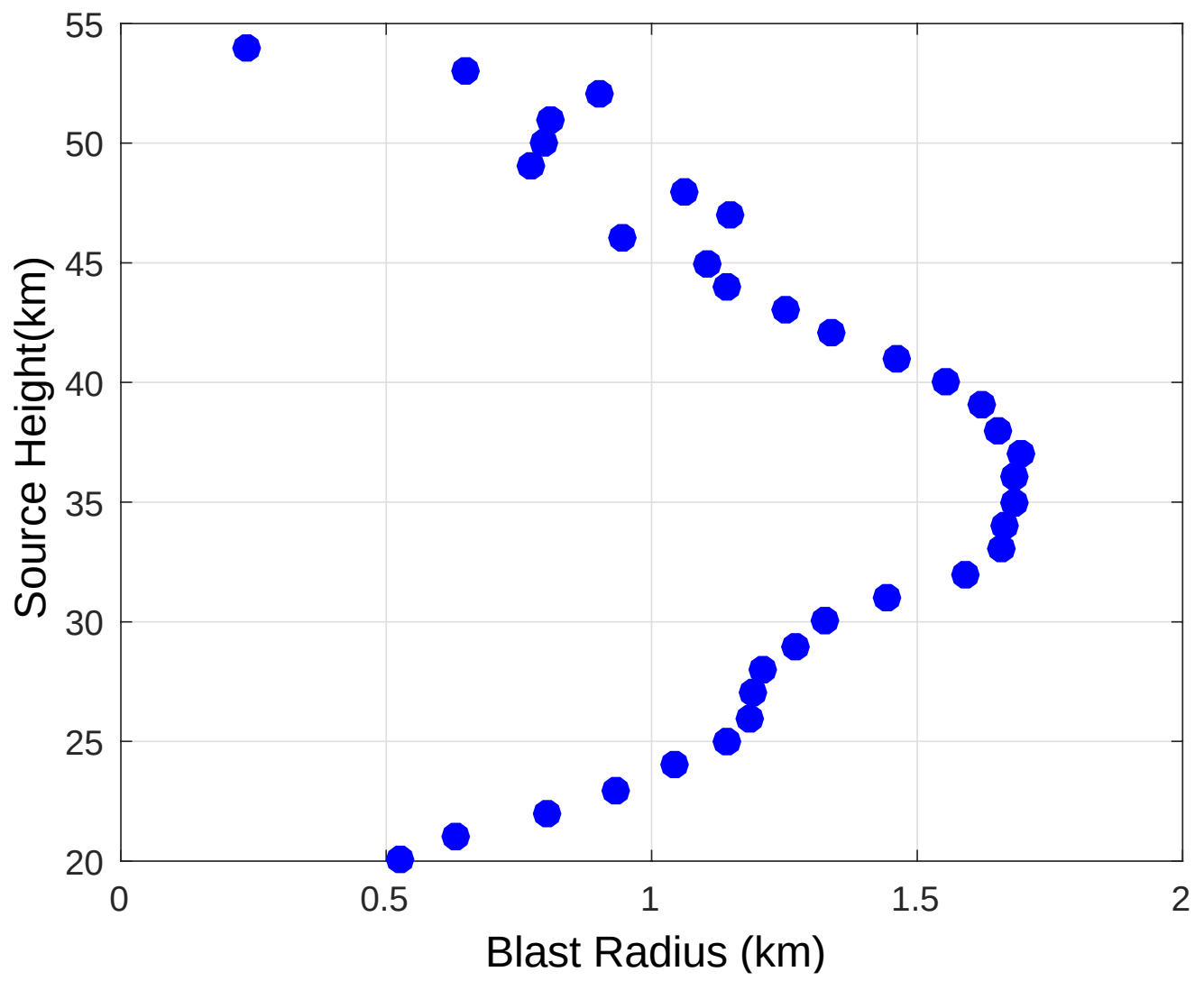

Fig 15: Blast radius plot for the September 3, 2004 Antarctica fireball.

As with the Chelyabinsk event, we applied raytracing to isolate the most probable source heights from each station and then weak shock modelling to check the predicted period against the observed period. We were not able to find any stratospherically ducted eigenrays for the furthest station, IS35, so we only have raytracing results for the first two closest stations, IS27 and IS55. The results are shown in Fig. 16a/b. For the IS27 composite plot, the elevation angle and back azimuth suggests that ~35km is the source height though these are not strongly constrained solutions. This height is also consistent with the ballistic angle closest to 90, consistent within uncertainty to a true ballistic arrival assuming some non-linear shock behaviour near the source (Brown et al., 2007). This is the most internally consistent height for a true cylindrical shock for this station. According to Brown et al., (2007) the angular deviation of ballistic angle can be up to 24 degrees. For this event, we see up to about 30 degrees deviation, which would not **be** unreasonable for such an energetic event where the shock is strongly non-linear for considerable distance from the trail and bending of the wave front may also be pronounced. In contrast, the IS55 composite plot, has a less well determined source height, with almost all heights showing one or two eigenray elevation arrivals agreeing with observations, while the back azimuth plot is uniformly at variance at all heights with observations. The ballistic arrival condition is met near 40-45 km height. As a whole, this does not suggest we can assign a unique source height to IS55 from raytracing alone.

a. IS27

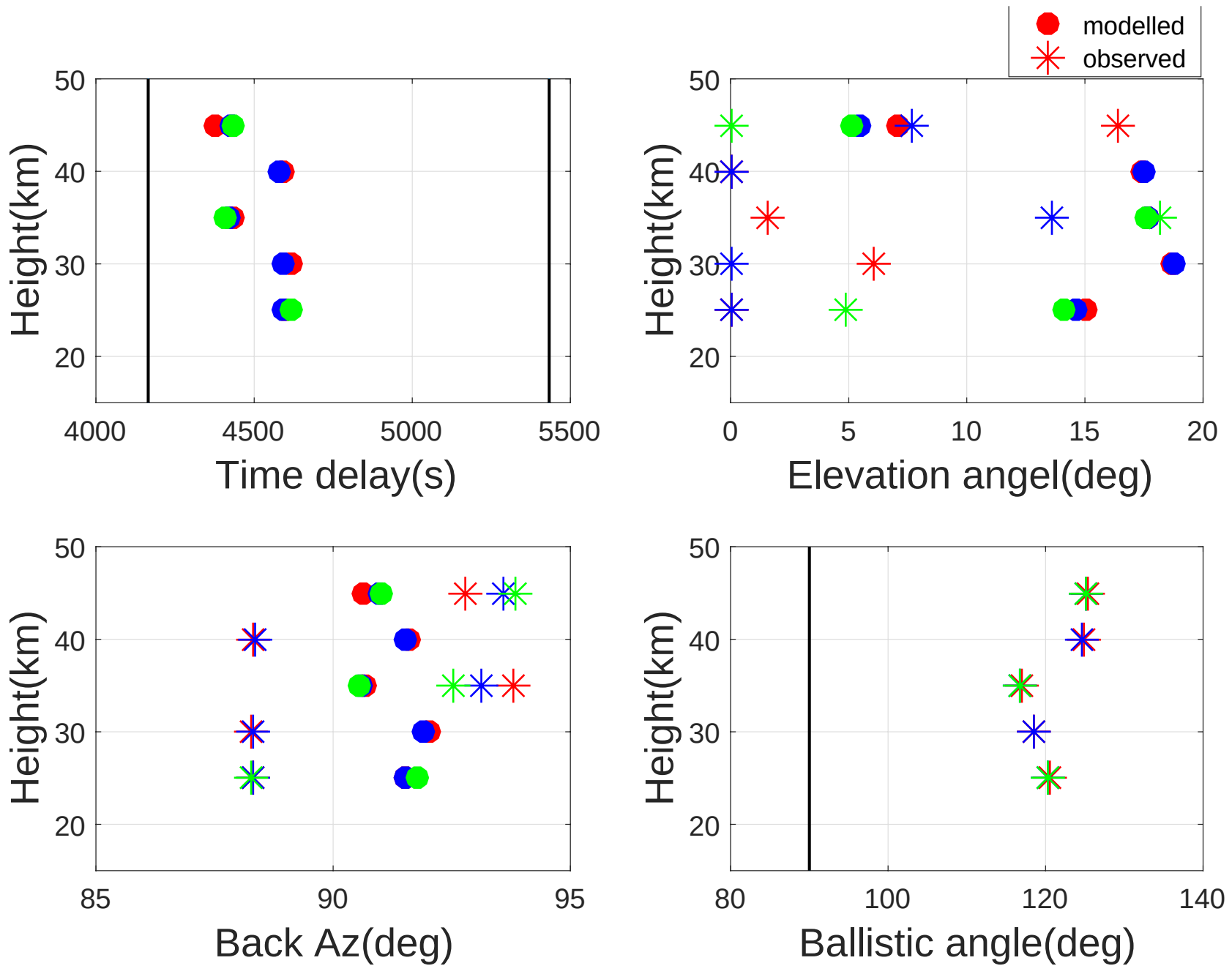

b. IS55

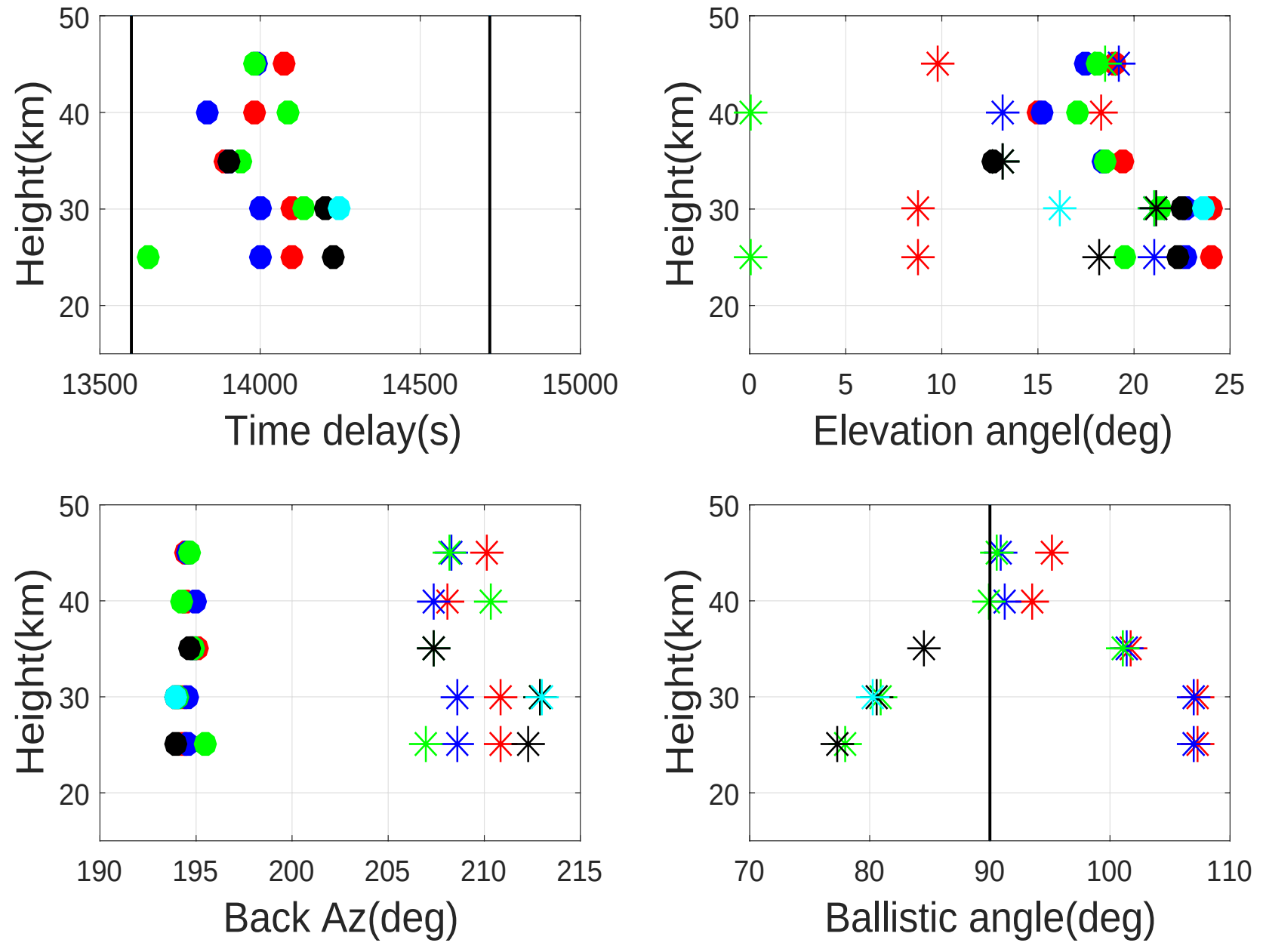

Fig 16: A composite plot showing the travel time with two vertical lines indicating where the signal starts and ends (top left), elevation angle at arrival times for the given eigenrays(top right), back azimuth at arrival for the given eigenrays (lower left) and ballistic angle of emission at the fireball trajectory (lower right) for eigenrays reaching each of IS27 (a) and IS55 (b) for the September 3, 2004, Antarctica fireball. The circles represent individual eigenrays from the modelled raytracing arrivals and asterisks are the observed quantity corresponding to the same colour circles at the modelled arrival times.

We next applied the ReVelle weak shock model to this event, converting each blast radii as a function of height into the equivalent period expected for the geometry to the observing station. The results from all three infrasound stations are shown in Fig. 17. For the closest station, I27, which also has the best estimate of source height (35 km), the observed and predicted signal period show good agreement for a source height of 35km. However, while the periods at other stations are consistent with high altitude sources, we cannot assign unique source height based on raytracing results so these are less convincing.

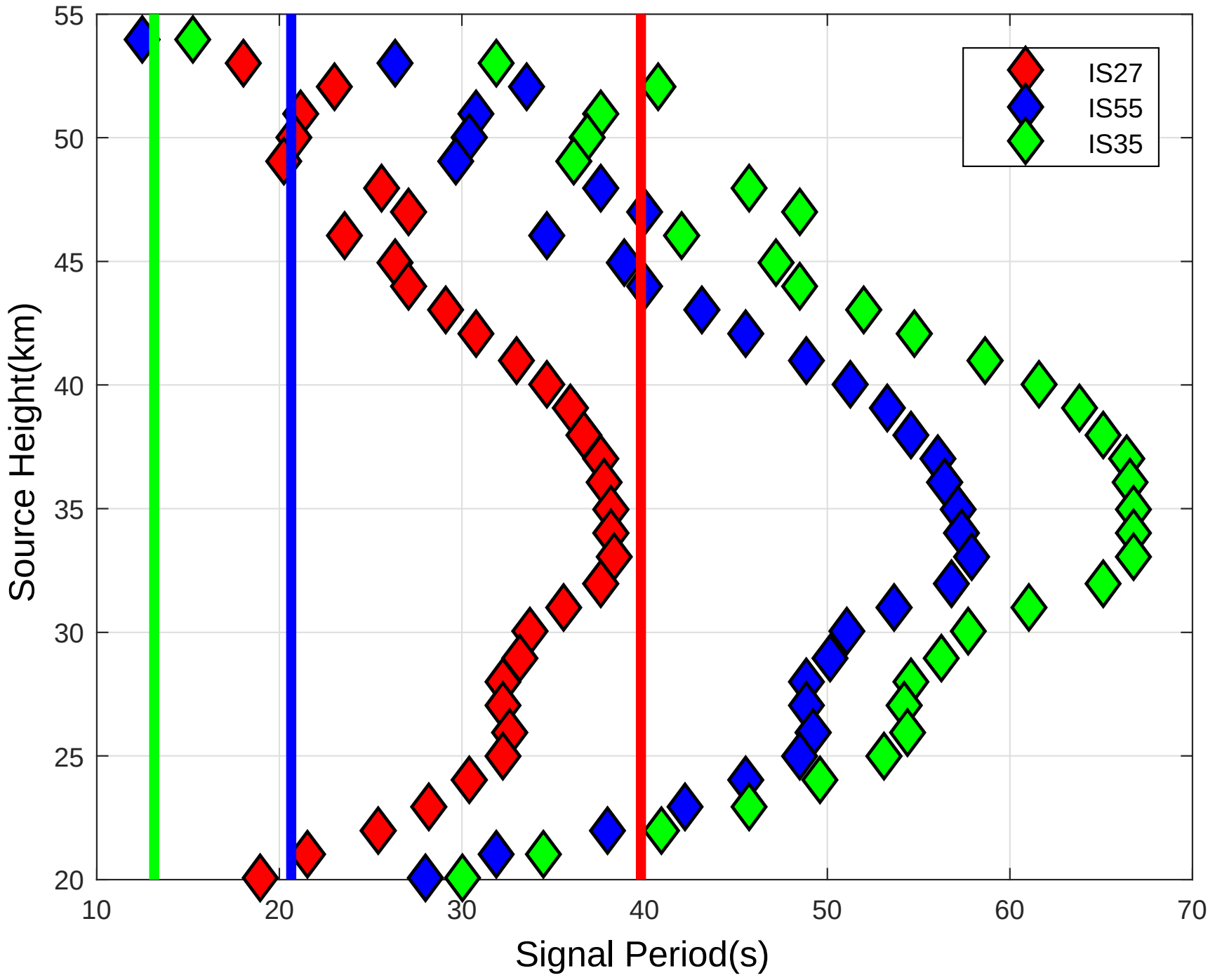

Fig 17: The signal period plot for the September 3, 2004, Antarctica fireball. The diamonds represent the simulated signal period from the weak shock model expected at the observing stations while the solid lines correspond to the observed infrasonic signal period (see Table S2 in Supplementary material). The different colors correspond to different infrasound stations with red, blue, green being IS27, IS55, IS35 respectively.

### 4.2.3 The Park Forest fireball – March 27, 2003

Our last case study event was Park Forest fireball, which occurred on March 27, 2003 at 5:50UT in Illinois, United States. The infrasound signal was detected at IS10 and at Blossom Point,

Maryland, however, in this paper, we will only be discussing the signal detected at the IS10 infrasound station, as the Blossom Point data is not publically available. According to Brown et al. (2004), the original total energy of this event was ~0.5 kT. To calculate the blast radii for the Park Forest event, we used same method as used to generate Fig. 13 for the Antarctica event. The blast radii graph can be found in Supplementary Material figure S1. Raytracing results (Fig. 18) did not clearly indicate the source height, however ballistic angle suggests range of 15-30km is the best source height presuming a cylindrical line source while the weak shock model (Fig. 19) is consistent with a height range of 20-25km. We can see that there is overlap agreement between these two predictions and therefore, we can conclude that a source height in the range of 20-25km is most probable.

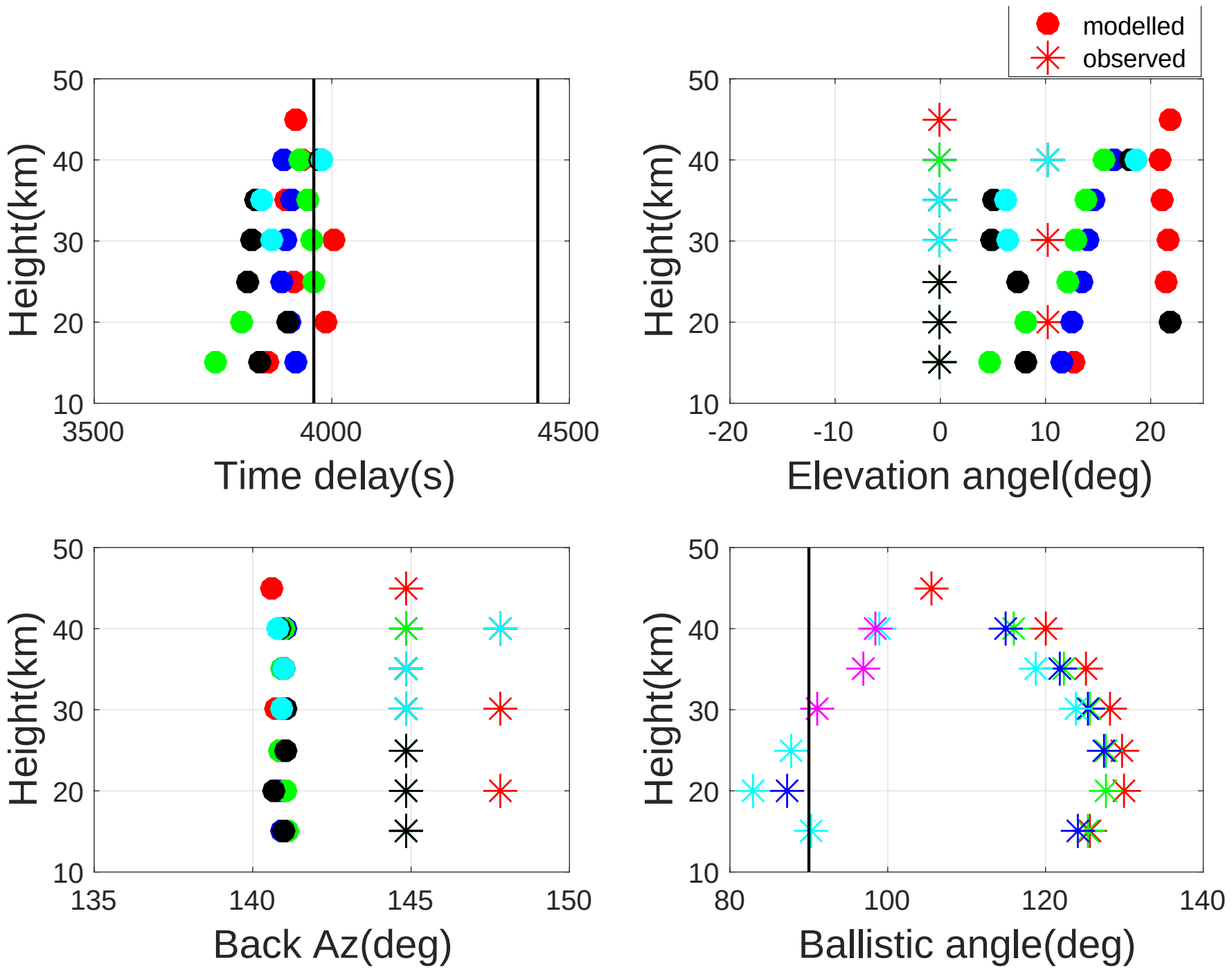

Fig 18: A composite plot for the Park Forest fireball infrasound showing the travel time with two vertical lines indicating where the signal starts and ends (top left), elevation angle at arrival (top right), back azimuth at arrival (lower left) and ballistic angle of emission at the fireball trajectory (lower right) for eigenrays reaching IS10. The circles represent individual eigenrays from the modelled raytracing arrivals and asterisks are the observed quantity corresponding to the same colour circles.

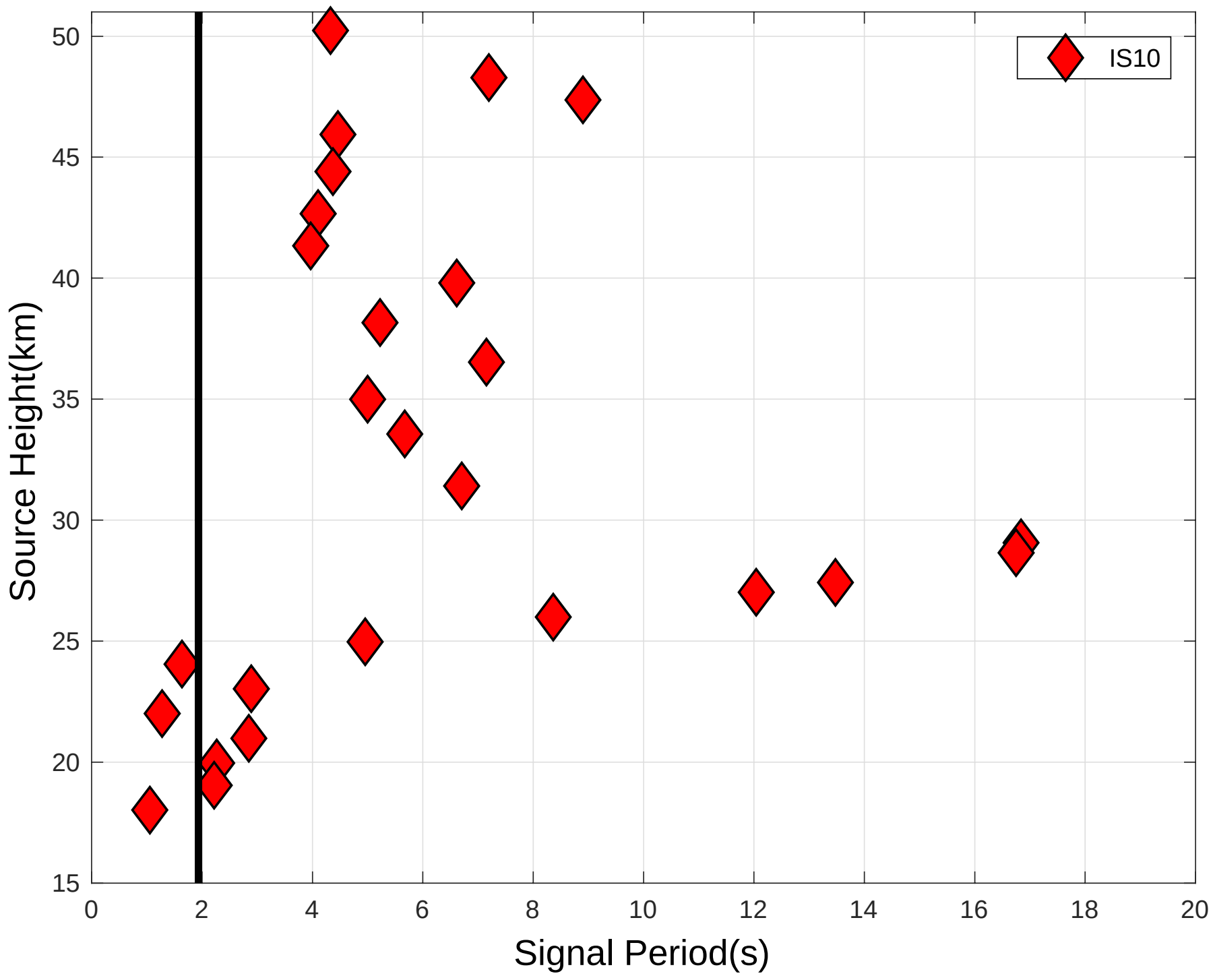

Fig 19: The signal period plot for the Park Forest meteorite dropping fireball of March 27, 2003. The diamonds represent simulated signal period from weak shock model and the solid line corresponds to the observed infrasonic signal period (see Table S2 in Supplementary Material).

## 5. Conclusions

In this paper, we extended the study of Edwards et al. (2006) and Ens et al. (2012) to examine the correlation between the infrasound signals and bolide characteristics, including entry angle, speed, height of peak brightness and range to station. Using a dataset consisting of 78 bolides detected by US government sensors we have analyzed 179 individual infrasonic waveforms and have been able to establish an empirical quantitative relationship between observed infrasonic bolide periods and total bolide yield (Eq. 9). Our period-yield relation for averaged signal periods was found to be very close to the AFTAC period-yield relation derived from nuclear tests, as was also found by Ens et al. (2012).

We find that two effects show a correlation with interstation periods:

1. Station noise levels produce noticeable scatter in period measurements, suggesting this may be a contributing cause to some of the large scatter, particularly for low SNR recordings. This is consistent with the results from Golden et al. (2012) who found that the dominant frequency in some cases occurs within a broader plateau making period measurements imprecise.
2. Increasing range from the bolide (particularly for larger events) shows a correlation with increasing apparent period. This is expected based on the larger attenuation with range with increasing frequency.

It is notable that the original AFTAC data nuclear yield-period data (Fig. 1) show significant scatter and that both of the foregoing effects would apply equally to the AFTAC or bolide datasets.

No empirical correlation with height of peak bolide brightness or entry angle is found for averaged signal periods. We suggest this implies either that the location along the trail where peak brightness occurs is not where the infrasonic periods dominantly originate, that each station sees a different part of the trail and/or the light curve for each event is quite different. The non-uniformity of energy release along the trail may cause a change in the shape of the shock wave during propagation through interaction of shocks formed at different parts of the trail. This is another possibility as to why we do not see a strong dependence of signal period with the source height. We applied the raytracing method and the ReVelle (1974) weak shock model to three fireball events to critically investigate how the bolide secondary characteristics, especially the source height affect the infrasound signal period. The main results of our case studies were:

- The weak shock model cannot be applied when the bolide blast radius is comparable to or larger than the atmospheric scale height (as is the case for Chelyabinsk).
- For relatively short-range stations (<1000km), heights from raytracing and the weak shock model were generally in good agreement. We found self-consistent results for a source height of ~35km for the measured infrasonic period at IS27 for the Antarctica event and 20-25km height for the infrasound period detected at IS10 for the Park Forest fireball.
- For longer-range stations, or stations with non-ballistic arrivals, we were not able to isolate a self-consistent and unique source height from raytrcing.

Our initial exploration is suggestive that source height may be at least part of the answer to station period spreads from bolide returns. However, the number of useable cases with sufficient information is too small to make any firm conclusions. More infrasonically detected bolides with complete energy deposition profiles (lightcurves) are required to test this hypothesis. It is clear, however, that much of the station period scatter is due to both station noise levels and range effects, probable explanations also for the scatter in AFTAC nuclear period measurements where differing source heights/locations are not an issue.

Finally, we note that the agreement between the bolide total yield – period relations and the AFTAC period-yield remains puzzling. Bolide yield from the US government sensor data represents all initial energy of the impactor at the top of the atmosphere. In contrast, the period observed at the ground probes only the energy deposition per unit trail length at some point (or points) along the trail, which should skew the bolide yield curve to lower apparent periods compared to the total yield. However, a countervailing factor is that the bolide trail segments occur at higher altitudes than the AFTAC nuclear detonations and this artificially increases the apparent period. The fact that the AFTAC energy-period and the bolide energy-period relations agree may simply be a reflection of the near balancing effects of burst height and energy deposition per unit length. In cases where the signals are detected at large distances from the trajectory, the source could be considered as point-like. If the source is point-like and occurring near ground level we would expect to see agreement with the AFTAC energy-period relations. The dataset used in this paper is provided in Table S1 together with all the raw measurements from Ens et al. (2012) in Table S2 in the Supplementary Material. In total this represents 128 individual bolides measured at 267 infrasound stations.

**Acknowledgements**
The authors thank Olga Popova and one anonymous reviewer for the detailed reviews of the original manuscript. This work was supported by grants from the Canadian Research Chair program, the Natural Sciences and Engineering Research council of Canada and Natural Resources Canada as well as NASA co-operative agreement NNX15AC94A.

**References**

Antolik, M., Ichinose, G., Creasey, J., & Clauter, D. (2014). Seismic and Infrasonic Analysis of the Major Bolide Event of 15 February 2013. *Seismological Research Letters*, 85(2), 334-343.

Bedard, A. J., & Georges, T. M. (2000). Atmospheric infrasound, *Physics Today, 53*(3), 32-37. doi:10.1063/1.883019.

Blom, P., 2014. GeoAc: Numerical Tools to Model Acoustic Propagation in the Geometric Limit, Software, Los Alamos National Laboratory.

Borovička, J., Spurný, P., & Brown, P. (2015). Small Near-Earth Asteroids as a Source of Meteorites. In P. Michel, F. E. DeMeo, & W. F. Bottke (Eds.), Asteroids IV (pp. 257-280). doi:10.2458/azu_uapress_9780816532131-ch014.

Borovička, J., Spurný, P., Brown, P., Wiegert, P., Kalenda, P., Clark, D., & Shrbený, L. (2013). The trajectory, structure and origin of the chelyabinsk asteroidal impactor. *Nature, 503*(7475), 235-237. doi:10.1038/nature12671.

Boslough, M. B., Brown, P., & Harris, A. W. (2015). Updated population and risk assessment for airbursts from near-Earth objects (NEOs). In: IEEE Aerospace Conference, Big Sky, Montana, (pp. 1-12).

Bowman, J. R., Baker, G. E., & Bahavar, M. (2005). Ambient infrasound noise. *Geophysical Research Letters, 32*, L09803. doi:10.1029/2005GL022486.

Brachet, N., Brown, D., Le Bras, R., Cansi, Y., Mialle, P., & Coyne, J. (2010). Monitoring the Earth's atmosphere with the global IMS infrasound Network. In Le Pichon, E. Blanc, & A. Hauchecorne (Eds.), Infrasound Monitoring for Atmospheric Studies, Chapter 3. © Springer Science. Business Media B.V, (pp. 77-118). doi: 10.1007/978-1-4020-9508-5_3.

Brown, P., Pack, D., Edwards, W. N., ReVelle, D. O., Yoo, B. B., Spalding, R. E., & Tagliaferri, E. (2004). The orbit, atmospheric dynamics, and initial mass of the Park Forest meteorite. *Meteoritics and Planetary Science, 39*(11), 1781-1796. doi:10.1111/j.1945 5100.2004.tb00075.x.

Brown, P. G., Assink, J. D., Astiz, L., Blaauw, R., Boslough, M. B., Borovicka, J., . . . Krzeminski, Z. (2013). A 500-kiloton airburst over chelyabinsk and an enhanced hazard from small impactors. *Nature, 503*(7475), 238.

Brown, P. G., Dube, K., & Silber, E. (2014). Detecting NEO Impacts using the International Monitoring System. In *AAS/Division for Planetary Sciences Meeting Abstracts* (p. 403.02). volume 46 of AAS/Division for Planetary Sciences Meeting Abstracts.

Brown, P. G., Edwards, W. N., ReVelle, D. O., & Spurny, P. (2007). Acoustic analysis of shock production by very high-altitude meteors—I: Infrasonic observations, dynamics and luminosity. *Journal of Atmospheric and Solar-Terrestrial Physics, 69*(4), 600-620. doi:10.1016/j.jastp.2006.10.011.

Brown, P. G., Spalding, R. E., ReVelle, D. O., Tagliaferri, E., & Worden, S. P. (2002). The flux of small near-Earth objects colliding with the Earth, *Nature, 420*(6913), 294–296. doi:10.1038/nature01238.

Brown, P. G., Wiegert, P., Clark, D., & Tagliaferri, E. (2015). Orbital and Physical Characteristics of Meter-sized Earth Impactors. In *AAS/Division for Planetary Sciences Meeting Abstracts* (p. 402.07). volume 47 of AAS/Division for Planetary Sciences Meeting Abstracts.

Cansi, Y. (1995). An automatic seismic event processing for detection and location: The PMCC method, *Geophysical Research Letters*, 22, 1021-1024, doi:10.1029/95GL00468.

Ceplecha, Z., Borovička, J., Elford, W. G., ReVelle, D.O.,Hawkes, R.L., Porubčan, V., & Šimek, M. (1998). Meteor phenomena and bodies. *Space Science Reviews*, *84*(3), 327–471. doi:10.1023/A:1005069928850.

de Groot-Hedlin, C. D., Hedlin, M. A. H., Drob, D. P. (2009). Atmospheric variability and infrasound monitoring. In A. Le Pichon, E. Blanc, & A. Hauchecorne (Eds.), Infrasound Monitoring for Atmospheric Studies, Chapter 15. © Springer Science. Business Media B.V, (pp. 475-507). doi:10.1007/978-1-4020-95085_15.

Drob, D. P., Emmert, J. T., Meriwether, J. W., Makela, J. J., Doornbos, E., Conde, M., Hernandez, G., Noto, J., Zawdie, K. A., McDonald, S. E., Huba, J. D., Klenzing, J. H. (2015). An update to the horizontal wind model (HWM): The quiet time thermosphere. *Earth and Space Science*, *2*(7), 301-319. doi:10.1002/2014EA000089.

Edwards, W. N. (2009). Meteor generated infrasound : Theory and observation. In A. Le Pichon, E. Blanc, & A. Hauchecorne (Eds.), Infrasound Monitoring for Atmospheric Studies, Chapter 12. © Springer Science. Business Media B.V,  (pp. 361-414). doi:10.1007/978-1-4020-9508 5_12.

Edwards, W. N., Brown, P. G., & ReVelle, D. O. (2005). Bolide Energy Estimates from Infrasonic measurements. *Earth, Moon and Planets*, *95*(1), 501-512. doi:10.1007/s11038-005-2244-4.

Edwards, W. N., Brown, P.G., & ReVelle, D.O. (2006). Estimates of meteoroid kinetic energies from observations of infrasonic airwaves. *Journal of Atmospheric and Solar-Terrestrial Physics*, *68*(10), 1136–1160. doi:10.1016/j.jastp.2006.02.010.

Ens, T. A., Brown, P. G., Edwards, W. N., & Silber, E. A. (2012) Infrasound production by bolides: A global statistical study. *Journal of Atmospheric and Solar-Terrestrial Physics*, *80*, 208-229. doi:10.1016/j.jastp.2012.01.018.

Few, A. A. (1969) Power spectrum of thunder, *Journal of Geophysical Research*, *74*, 6926-6934, doi:10.1029/JC074i028p06926.

Golden, P., Negraru, P., & Howard, J. (2012). Infrasound Studies for Yield Estimation of HE Explosions. AFRL-RV-PS-TR-2012-0084. Final report. June 5th 2012.

Harris, M., & Young, C. (1997). MatSeis: a seismic GUI and tool-box for MATLAB. *Seismological Research Letters*, *68*(2), 267-269.

Herrin, G., Bass, H., Andre, B., Woodward, B., Drob, D., Hedlin, M., Garcs, M., Golden, P., Norris, D., de Groot-Hedlin, C., Walker, K., Szuberla, C., Whitaker, R., & Shields, D. (2008). High-altitude infrasound calibration experiments. In D. Stern (Ed.), *Acoustic Today* (pp. 9-21). volume 4. doi:10.1121/1.2961169.

Klekociuk, A. R., Brown, P. G., Pack, D. W., ReVelle, D. O., Edwards, W. N., Spalding, R. E., Tagliaferri, E., Yoo, B. B., & Zagari, J. (2005). Meteoritic dust from the atmospheric disintegration of a large meteoroid. *Nature*, *436*, 1132-1135. doi:10.1038/nature03881.

Norris, D., Gibson, R., & Bongiovanni, K. (2010). Numerical methods to model infrasonic propagation through realistic specifications of the atmosphere. . In A. Le Pichon, E. Blanc, & A. Hauchecorne (Eds.), Infrasound Monitoring for Atmospheric Studies, Chapter

17. © Springer Science. Business Media B.V, (pp. 541-574). doi:10.1007/978-1-4020-9508 5_17.
Picone, J.M., Hedin, A.E., Drob, D.P., & Aikin, A.C. (2002). NRLMSISE- 00 empirical model of
the atmosphere: Statistical comparisons and scientific issues. *Journal of Geophysical Research: Space Physics*. (1978–2012), 107(A12), SIA 15-1-SIA 15-16. doi:10.1029/2002JA009430.
Pilger, C., Ceranna, L., Ross, J. O., Le Pichon, A., Mialle, P., & Garcés, M. A. (2015) CTBT infrasound network performance to detect the 2013 Russian fireball event. *Geophysics Research Letter, 42*(7), 2523–2531.
Popova, O. P., Jenniskens, P., Emel'yanenko, V., Kartashova, A., Biryukov, E., Khaibrakhmanov, S., . . . the Chelyabinsk Airburst Consortium. (2013). Chelyabinsk airburst, damage assessment, meteorite recovery, and characterization. *Science, 342*(6162), 1069-1073. doi:10.1126/science.1242642.
ReVelle, D. O. (1974). Acoustics of meteors-effects of the atmospheric temperature and wind structure on the sounds produced by meteors. Ph.D. Dissertation, University of Michigan, Ann Arbor, MI, USA.
ReVelle, D. O. (1976). On meteor generated infrasound, *Journal of Geophysical Research, 81*(7), 1217-1230, doi: 10.1029/JA081i007p01217.
ReVelle, D. O. (1997). Historical Detection of Atmospheric Impacts by Large Bolides Using Acoustic-Gravity Waves. *Annals of the New York Academy of Sciences, 822*(1 Near-Earth Ob), 284-302, doi: 10.1111/j.1749-6632.1997.tb48347.x.
Richard P. K. & Timothy S. M. (2000). Infrasound sensor models and evaluation, Sandia National Laboratories.
Silber, E. A., & Brown, P. G. (2014). Optical observations of meteors generating infrasound—I: Acoustic signal identification and phenomenology. *Journal of Atmospheric and Solar-Terrestrial Physics, 119*, 116-128. doi:10.1016/j.jastp.2014.07.005
Silber, E. A., Brown, P. G., & Krzeminski, Z. (2015). Optical observations of meteors generating infrasound: Weak shock theory and validation. *Journal of Geophysical Research (Planets), 120*, 413-428. doi:10.1002/2014JE004680.arXiv:1411.5406.
Silber, E. A., Le Pichon, A., & Brown, P. G. (2011). Infrasonic detection of a near-Earth object impact over Indonesia on 8 October 2009, *Geophysical Research Letters*, *38*(12), L12201. doi:10.1029/2011GL047633.
Silber, E. A., ReVelle, D. O., Brown, P. G., & Edwards, W. N. (2009). An estimate of the terrestrial influx of large meteoroids from infrasonic measurements. *Journal of Geophysical Research (Planets)*, *114*(E8), E08006. doi:10.1029/2009JE003334.
Stevens, J. L., Divnov, I. I., Adams, D. A., Murphy, J. R., & Bourchik, V. N. (2002). Constraints on infrasound scaling and attenuation relations from soviet explosion data. *Pure and Applied Geophysics, 159*(5), 1045-1062. doi:10.1007/s00024-002-8672-4.
Towne, D.H. (1967) Wave Phenomena, Reading, Addison-Wesley Pub. Co., Massachusetts.
Tsikulin, M. A. (1970). Shock waves during the movement of large meteorites in the atmosphere (No. NIC-Trans-3148). Naval Intelligence Command Alexandria VA Translation Div.
Young, C. J., Chael, E. P., & Merchant, B. J. (2002). Version 1.7 of MatSeis and the GNEM R&E regional seismic analysis tools. Proceedings, 24th Seismic Research Review. (pp. 915–924).

**Supplementary Material**

Table S1: Summary of infrasound signal characteristics from 78 individual bolide events detected at 179 infrasound stations between 2006 to 2015 following the bolide infrasound analysis of Edwards et al., (2006) and Ens et al., (2012). The table shows for each detection; bolide date/time from the JPL webpage, infrasound station, signal arrival time, signal duration, range from the bolide location to station, theoretical back azimuth (Az), observed back azimuth (Az), JPL energy measured by U.S. government sensors, peak-to-peak (P2P) amplitude, period at maximum amplitude computed using the zero crossings method (Edwards et al., 2006), period from inversion of frequency at the maximum power spectral density (PSD), bolide integrated energy signal to noise ratio (SNR) and bandpass used for measurements. The highlighted events are common with Ens et al. (2012).

| Date/ [Time] | Station | Arrival time (UT) | Duration (s) | Range (km) | Theo. Az (deg) | Obs. Az (deg) | JPL Energy (kT) | P2P Amp (Pa) | Period @Max Amp (s) | Period @Max PSD (s) | Bolide Integrated Energy SNR | Band-pass (Hz) |
|---|---|---|---|---|---|---|---|---|---|---|---|---|
| **19-Feb-16** [08:15:02] | I17CI | 09:10:08 | 356 | 1000 | 229 | 230 | 0.56 | 0.15 | 3.58 | 4.40 | 3.33 | 0.08-2.5 |
| | I50GB | 09:08:03 | 439 | 1009 | 18 | 14 | 0.56 | 0.39 | 2.19 | 3.26 | 2.37 | 0.5-4 |
| **06-Feb-16** [13:55:09] | I49GB | 15:23:34 | 120 | 1425 | 298 | 296 | 13 | 0.11 | 5.87 | 7.31 | 1.11 | 0.25-1.5 |
| | I27DE | 18:06:14 | 360 | 4601 | 337 | 334 | 13 | 1.50 | 6.61 | 6.94 | 33.08 | 0.02-3.5 |
| | I11CV | 18:51:18 | 142 | 5083 | 183 | 158 | 13 | 0.06 | 2.42 | 2.35 | 1.44 | 0.4-1.2 |
| **21-Dec-15** [02:32:48] | I39PW | 03:28:07 | 519 | 951 | 101 | 100 | 0.26 | 0.28 | 3.18 | 3.71 | 1.75 | 0.2-4 |
| **13-Oct-15** [12:23:08] | I08BO | 14:16:08 | 160 | 1959 | 64 | 41 | 0.08 | 0.03 | 1.60 | 1.45 | 0.65 | 0.4-2 |
| **08-Sep-15** [13:46:42] | I32KE | 14:49:15 | 259 | 1138 | 318 | 315 | 0.07 | 0.06 | 1.64 | 1.51 | 7.32 | 0.5-2.5 |
| **07-Sep-15** [01:41:19] | I45RU | 06:08:06 | 174 | 4545 | 234 | 236 | 3.9 | 0.01 | 1.99 | 1.69 | 1.12 | 0.5-1.5 |
| | I46RU | 05:54:31 | 505 | 4555 | 159 | 160 | 3.9 | 0.04 | 3.34 | 3.44 | 2.65 | 0.2-1.2 |
| | I04AU | 07:03:19 | 672 | 5762 | 338 | 336 | 3.9 | 0.05 | 1.15 | 1.07 | 2.87 | 0.8-2 |
| | I53US | 10:42:44 | 534 | 9596 | 297 | 300 | 3.9 | 0.02 | 2.97 | 2.80 | 2.80 | 0.2-1.5 |
| **2-Sep-15** [20:10:30] | I31KZ | 21:53:55 | 463 | 1879 | 233 | 232 | 0.13 | 0.06 | 1.71 | 1.87 | 15.25 | 0.3-4 |
| | I48TN | 22:45:14 | 336 | 2736 | 73 | 72 | 0.13 | 0.01 | 1.72 | 1.86 | 2.17 | 0.5-1.5 |
| | I46RU | 23:36:04 | 323 | 3712 | 262 | 257 | 0.13 | 0.02 | 2.67 | 2.59 | 3.57 | 0.25-1.5 |
| **14-Jun-15** [03:03:06] | I39PW | 04:11:55 | 109 | 1161 | 264 | 264 | 0.22 | 0.01 | 0.91 | 0.87 | 1.74 | 1-3 |
| **10-May-15** [07:45:01] | I36NZ | 08:02:31 | 147 | 345 | 219 | 222 | 0.49 | 0.81 | 1.74 | 2.05 | 9.62 | 0.45-7 |
| **30-Apr-15** [10:21:01] | I13CL | 13:23:44 | 138 | 3503 | 219 | 217 | 0.32 | 0.09 | 0.63 | 0.75 | 1.02 | 0.7-2.1 |
| **08-Apr-15** [04:06:31] | I33MG | 04:33:22 | 339 | 841 | 110 | 109 | 0.49 | 0.63 | 2.01 | 2.04 | 59.26 | 0.4-3 |
| **11-Mar-15** [06:18:59] | I39PW | 07:45:04 | 888 | 1703 | 273 | 272 | 0.23 | 0.25 | 2.68 | 2.58 | 19.28 | 0.3-6 |
| **04-Mar-15** [04:30:05] | I06AU | 05:25:47 | 519 | 1029 | 245 | 244 | 0.18 | 0.23 | 2.39 | 2.61 | 2.48 | 0.3-3 |
| | I52GB | 06:15:11 | 357 | 1945 | 121 | 122 | 0.18 | 0.10 | 1.79 | 1.89 | 4.37 | 0.5-3 |
| **26-Feb-15** [22:06:24] | I53US | 22:24:46 | 152 | 351 | 352 | 358 | 0.53 | 1.86 | 3.90 | 3.79 | 27.78 | 0.1-6 |
| | I18DK | 00:23:42 | 720 | 2561 | 290 | 282 | 0.53 | 0.43 | 3.91 | 3.88 | 13.76 | 0.2-4 |
| | I56US | 00:38:44 | 50 | 2820 | 333 | 329 | 0.53 | 0.03 | 1.15 | 1.31 | 4.42 | 0.5-1.3 |
| | I10CA | 01:05:51 | 398 | 3452 | 325 | 324 | 0.53 | 0.14 | 3.14 | 3.62 | 9.25 | 0.2-2 |
| **09-Jan-15** [10:41:11] | I32KE | 11:34:16 | 480 | 963 | 292 | 291 | 0.41 | 0.08 | 1.16 | 1.18 | 5.01 | 0.5-4 |
| | I19DJ | 12:36:20 | 132 | 1903 | 237 | 237 | 0.41 | 0.08 | 2.01 | 1.99 | 1.75 | 0.45-1.2 |
| **07-Jan-15** [01:05:59] | I26DE | 02:09:25 | 412 | 1053 | 104 | 105 | 0.4 | 0.32 | 2.85 | 2.79 | 14.31 | 0.3-6 |
| | I43RU | 02:22:15 | 888 | 1417 | 215 | 214 | 0.4 | 0.18 | 3.24 | 2.95 | 2.49 | 0.3-4 |
| **13-Dec-14** [02:53:52] | I18DK | 04:15:28 | 201 | 1428 | 345 | 337 | 0.15 | 0.08 | 1.68 | 1.81 | 2.70 | 0.4-2 |
| | I53US | 05:11:39 | 631 | 2440 | 358 | 5 | 0.15 | 0.04 | 1.32 | 1.12 | 4.42 | 0.7-3 |
| **12-Dec-14** [06:48:11] | I30JP | 07:15:02 | 227 | 466 | 114 | 108 | 0.11 | 0.11 | 1.56 | 1.31 | 4.30 | 0.7-3 |
| | I45RU | 08:23:05 | 425 | 1629 | 133 | 128 | 0.11 | 0.07 | 2.08 | 2.06 | 2.18 | 0.4-1.5 |
| **28-Nov-14** [11:47:18] | I36NZ | 12:07:02 | 349 | 364 | 126 | 130 | 1.7 | 4.06 | 2.74 | 2.75 | 44.19 | 0.01-9 |
| | I05AU | 14:34:08 | 988 | 3150 | 111 | 112 | 1.7 | 0.24 | 9.69 | 10.11 | 3.32 | 0.05-4 |
| | I22FR | 14:48:58 | 749 | 3212 | 150 | 150 | 1.7 | 0.12 | 2.85 | 3.03 | 2.57 | 0.25-1.5 |
| | I55US | 15:06:37 | 270 | 3654 | 26 | 337 | 1.7 | 0.03 | 3.00 | 2.79 | 4.87 | 0.3-2 |
| | I07AU | 17:02:22 | 265 | 5590 | 134 | 133 | 1.7 | 0.07 | 3.41 | 3.36 | 1.79 | 0.25-1.1 |
| **26-Nov-14** [17:40:16] | I27DE | 18:16:24 | 129 | 668 | 287 | 285 | 0.32 | 0.04 | 1.62 | 1.74 | 10.62 | 0.5-7.5 |
| **04-Nov-14** [20:13:30] | I45RU | 21:17:01 | 631 | 1305 | 270 | 272 | 0.45 | 0.30 | 2.66 | 2.90 | 6.54 | 0.2-1 |
| **14-Oct-14** [10:25:03] | I39PW | 12:48:17 | 590 | 2006 | 239 | 318 | 0.1 | 0.03 | 1.25 | 1.52 | 1.54 | 0.6-2.5 |
| **23-Aug-14** [06:29:41] | I05AU | 08:42:45 | 463 | 2358 | 200 | 193 | 7.6 | 0.21 | 2.57 | 2.58 | 5.70 | 0.35-4 |
| **16-May-14** [12:42:48] | I36NZ | 12:45:17 | 120 | 39 | 144 | 135 | 0.82 | 5.49 | 3.23 | 3.72 | 155.22 | 0.1-3 |
| **08-May-14** [19:42:37] | I04AU | 21:57:32 | 401 | 2625 | 256 | 255 | 2.4 | 0.20 | 3.00 | 2.71 | 1.67 | 0.3-1.5 |
| **29-Mar-14** [13:45:41] | I07AU | 15:15:05 | 490 | 1623 | 231 | 230 | 0.13 | 0.05 | 1.90 | 1.62 | 4.83 | 0.5-2 |
| | I05AU | 16:20:48 | 352 | 2802 | 295 | 294 | 0.13 | 0.02 | 1.89 | 1.63 | 2.61 | 0.5-1.8 |
| | I06AU | 16:34:18 | 778 | 3150 | 130 | 129 | 0.13 | 0.03 | 1.96 | 1.60 | 1.76 | 0.6-3 |
| **12-Jan-14** [16:00:48] | I19DJ | 18:20:49 | 821 | 2526 | 111 | 113 | 0.24 | 0.12 | 3.42 | 4.01 | 3.87 | 0.2-6 |
| | I33MG | 18:49:41 | 141 | 3071 | 39 | 34 | 0.24 | 0.07 | 1.82 | 1.68 | 1.51 | 0.4-1.2 |
| | I32KE | 18:46:00 | 1059 | 3099 | 81 | 84 | 0.24 | 0.13 | 1.76 | 1.96 | 3.33 | 0.4-5 |
| | I47ZA | 21:09:43 | 369 | 5453 | 57 | 57 | 0.24 | 0.02 | 1.78 | 1.47 | 1.11 | 0.6-4 |
| **08-Jan-14** [17:05:34] | I39PW | 18:39:41 | 427 | 1750 | 124 | 124 | 0.11 | 0.04 | 1.77 | 1.91 | 1.82 | 0.5-1.5 |
| | I07AU | 19:25:07 | 276 | 2525 | 37 | 34 | 0.11 | 0.07 | 1.46 | 2.44 | 1.34 | 0.4-2.5 |
| **23-Dec-13** [08:30:57] | I48TN | 08:56:43 | 309 | 764 | 305 | 308 | 0.43 | 0.34 | 1.41 | 1.58 | 3.73 | 0.45-7 |
| | I26DE | 09:49:37 | 239 | 1394 | 226 | 214 | 0.43 | 0.24 | 2.31 | 2.17 | 5.14 | 0.4-2 |
| | I31KZ | 12:31:58 | 176 | 4459 | 277 | 276 | 0.43 | 0.03 | 1.73 | 1.93 | 1.94 | 0.3-1 |
| **08-Dec-13** [03:10:09] | I59US | 04:47:17 | 258 | 1730 | 330 | 333 | 0.2 | 0.04 | 1.68 | 1.54 | 2.47 | 0.6-2 |
| **21-Nov-13** [01:50:35] | I43RU | 03:05:28 | 262 | 1343 | 187 | 175 | 0.23 | 0.13 | 2.03 | 2.19 | 1.30 | 0.3-2.5 |
| | I31KZ | 03:35:04 | 72 | 1812 | 258 | 256 | 0.23 | 0.06 | 1.65 | 2.05 | 2.32 | 0.25-3 |
| **12-Oct-13** [16:06:45] | I50GB | 17:33:37 | 992 | 1690 | 222 | 221 | 3.5 | 0.65 | 4.81 | 5.98 | 4.66 | 0.1-5 |
| | I17CI | 19:26:10 | 820 | 3617 | 217 | 218 | 3.5 | 0.35 | 5.92 | 7.77 | 10.36 | 0.05-1.8 |
| | I35NA | 20:15:49 | 498 | 4461 | 263 | 253 | 3.5 | 0.06 | 10.44 | 8.81 | 1.42 | 0.05-1 |
| | I32KE | 22:48:03 | 764 | 7011 | 249 | 250 | 3.5 | 0.15 | 6.76 | 6.35 | 4.06 | 0.05-2 |
| **31-Jul-13** [03:50:14] | I05AU | 05:15:23 | 354 | 1512 | 318 | 323 | 0.22 | 0.11 | 1.52 | 1.85 | 6.02 | 0.5-4 |
| **27-Jul-13** [08:30:36] | I39PW | 10:52:51 | 544 | 2532 | 108 | 103 | 0.36 | 0.10 | 3.24 | 3.01 | 1.83 | 0.3-2 |
| | I22FR | 11:03:19 | 360 | 2754 | 334 | 336 | 0.36 | 0.09 | 1.98 | 2.00 | 4.16 | 0.4-2 |
| **30-Apr-13** [08:40:38] | I42PT | 09:05:32 | 486 | 460 | 212 | 216 | 10 | 1.27 | 6.77 | 8.03 | 10.89 | 0.2-8 |
| **21-Apr-13** [06:23:12] | I41PY | 07:00:42 | 600 | 747 | 253 | 251 | 2.5 | 3.21 | 6.16 | 5.06 | 32.84 | 0.07-8 |
| | I08BO | 07:41:05 | 163 | 1380 | 164 | 163 | 2.5 | 0.01 | 0.75 | 0.98 | 1.44 | 1-3 |
| | I09BR | 08:20:53 | 1076 | 2198 | 228 | 226 | 2.5 | 0.90 | 4.91 | 4.88 | 11.92 | 0.04-4 |
| | I02AR | 09:10:49 | 276 | 2953 | 5 | 4 | 2.5 | 0.05 | 1.79 | 1.16 | 2.20 | 0.6-3 |
| | I11CV | 12:32:59 | 469 | 6566 | 223 | 226 | 2.5 | 0.08 | 4.12 | 4.71 | 1.29 | 0.3-2 |
| **15-Feb-13** [03:20:33] | I31KZ | 03:48:08 | 1026 | 530 | 22 | 29 | 440 | 12.24 | 37.99 | 45.51 | 76.90 | 0.01-4 |
| | I43RU | 05:02:09 | 2142 | 1502 | 88 | 97 | 440 | 1.58 | 38.48 | 30.06 | 13.67 | 0.01-3 |
| | I46RU | 04:44:29 | 701 | 1532 | 283 | 268 | 440 | 2.52 | 21.08 | 16.72 | 35.76 | 0.02-4 |
| | I34MN | 06:14:55 | 924 | 3185 | 301 | 230 | 440 | 0.50 | 22.79 | 30.34 | 4.40 | 0.03-2 |
| | I26DE | 07:10:31 | 812 | 3257 | 60 | 56 | 440 | 0.97 | 28.98 | 29.26 | 2.89 | 0.03-3 |
| | I18DK | 08:17:13 | 1948 | 4893 | 39 | 17 | 440 | 2.73 | 49.99 | 69.72 | 13.75 | 0.01-3 |
| | I45RU | 07:55:27 | 887 | 5022 | 310 | 305 | 440 | 1.34 | 38.45 | 21.01 | 1.07 | 0.02-3 |
| | I44RU | **08:50:55** | **1493** | **5798** | **314** | **303** | **440** | **0.88** | **19.25** | **39.00** | **4.20** | **0.02-3** |
| | I53US | 09:36:30 | 1660 | 6481 | 341 | 339 | 440 | 12.62 | 25.69 | 19.74 | 165.14 | 0.02-4 |
| | I10CA | 11:29:57 | 1107 | 8147 | 14 | 355 | 440 | 2.75 | 32.53 | 36.41 | 5.55 | 0.01-1 |
| | I33MG | 11:20:35 | 720 | 8311 | 8 | 12 | 440 | 1.94 | 47.96 | 43.12 | 2.07 | 0.01-0.2 |
| | I56US | 11:35:08 | 1621 | 8554 | 1 | 352 | 440 | 1.59 | 29.45 | 30.91 | 7.61 | 0.015-3 |
| | I57US | 13:10:00 | 2000 | 10182 | 1 | 2 | 440 | 1.59 | 28.45 | 27.31 | 3.42 | 0.01-1 |
| | I59US | 13:42:44 | 377 | 11030 | 339 | 316 | 440 | 0.39 | 84.70 | 81.92 | 2.48 | 0.01-0.1 |
| | I27DE | 17:50:43 | 830 | 14983 | 49 | 61 | 440 | 1.86 | 45.15 | 44.28 | 3.76 | 0.01-0.1 |
| **20-Nov-12** [20:37:31] | I32KE | 21:24:19 | 452 | 905 | 297 | 299 | 0.09 | 0.18 | 1.24 | 1.11 | 43.34 | 0.2-6 |
| | I19DJ | 22:14:13 | 391 | 1799 | 237 | 238 | 0.09 | 0.01 | 2.37 | 2.28 | 3.21 | 0.4-2.5 |
| **18-Sep-12** [19:34:39] | I09BR | 21:21:14 | 388 | 1928 | 346 | 348 | 0.67 | 0.02 | 2.75 | 2.53 | 2.43 | 0.3-1.5 |
| | I08BO | 22:03:23 | 115 | 2633 | 44 | 42 | 0.67 | 0.03 | 0.90 | 0.91 | 2.04 | 1-1.8 |
| **10-Sep-12** [01:03:32] | I02AR | 03:34:48 | 435 | 2757 | 215 | 215 | 0.08 | 0.09 | 2.23 | 1.01 | 1.52 | 0.5-2.5 |
| **26-Aug-12** [14:55:47] | I39PW | 16:55:05 | 270 | 1980 | 285 | 282 | 0.68 | 0.02 | 1.40 | 1.11 | 2.44 | 0.6-2.5 |
| **12-Mar-12** [06:40:44] | I39PW | 07:25:22 | 86 | 807 | 134 | 136 | 0.3 | 0.05 | 1.71 | 1.44 | 1.32 | 0.5-2 |
| | I30JP | 09:58:59 | 140 | 3648 | 181 | 238 | 0.3 | 0.42 | 2.03 | 1.81 | 37.04 | 0.5-3 |
| **04-Feb-12** [14:42:51] | I48TN | 15:33:22 | 95 | 929 | 249 | 248 | 0.43 | 0.07 | 1.43 | 1.38 | 1.08 | 0.6-5.5 |
| **15-Jan-12** [12:26:20] | I55US | 14:39:53 | 207 | 2422 | 276 | 272 | 0.08 | 0.01 | 0.92 | 1.02 | 1.51 | 0.9-3 |
| **25-May-11** [05:40:02] | I35NA | 08:05:51 | 444 | 2619 | 351 | 353 | 4.8 | 0.08 | 2.31 | 1.78 | 1.72 | 0.4-2 |
| **06-Apr-11** [08:30:55] | I18DK | 09:24:54 | 593 | 1051 | 121 | 116 | 0.43 | 0.45 | 3.15 | 3.90 | 9.57 | 0.3-3 |
| **01-Mar-11** [10:37:54] | I34MN | 11:16:27 | 186 | 658 | 345 | 315 | 0.13 | 0.10 | 2.48 | 2.77 | 2.47 | 0.3-1 |
| **21-Feb-11** [05:07:03] | I31KZ | 07:45:50 | 562 | 2946 | 210 | 207 | 0.13 | 0.31 | 3.04 | 3.28 | 27.48 | 0.2-6 |
| **25-Dec-10** [23:24:00] | I30JP | 00:59:24 | 560 | 1603 | 74 | 72 | 33 | 0.59 | 6.61 | 6.88 | 0.79 | 0.2-2 |
| | I44RU | 00:58:16 | 544 | 1680 | 179 | 178 | 33 | 0.61 | 2.76 | 3.14 | 4.82 | 0.3-3 |
| | I45RU | 01:33:35 | 610 | 2276 | 99 | 93 | 33 | 0.74 | 3.37 | 3.26 | 8.06 | 0.25-3 |
| | I39PW | 03:39:50 | 649 | 4127 | 31 | 36 | 33 | 2.32 | 12.49 | 11.70 | 1.88 | 0.06-1 |
| | I34MN | 03:28:44 | 769 | 4257 | 85 | 15 | 33 | 0.02 | 1.90 | 2.34 | 1.09 | 0.4-3 |
| | I53US | 03:41:02 | 594 | 4572 | 256 | 256 | 33 | 2.02 | 8.35 | 8.19 | 35.92 | 0.1-3 |
| | I56US | 05:46:23 | 309 | 6625 | 295 | 295 | 33 | 0.78 | 15.63 | 15.17 | 3.43 | 0.01-1.3 |
| | I18DK | 05:40:26 | 1075 | 6758 | 318 | 308 | 33 | 0.49 | 11.32 | 11.38 | 7.04 | 0.07-2 |
| | I10CA | 06:41:36 | 677 | 7837 | 307 | 308 | 33 | 0.52 | 9.88 | 9.75 | 3.43 | 0.1-1.5 |
| | I55US | 11:33:45 | 737 | 12885 | 352 | 347 | 33 | 0.13 | 18.44 | 15.46 | 4.54 | 0.04-1 |
| | I08BO | 13:48:06 | 212 | 14887 | 308 | 156 | 33 | 0.10 | 7.91 | 7.06 | 1.11 | 0.1-0.8 |
| **03-Sep-10** [12:04:58] | I55US | 13:59:19 | 562 | 2005 | 326 | 339 | 3.8 | 0.42 | 7.54 | 7.73 | 7.54 | 0.1-6 |
| | I05AU | 14:04:10 | 202 | 2059 | 182 | 184 | 3.8 | 0.07 | 1.86 | 1.88 | 1.49 | 0.5-2.5 |
| | I04AU | 15:55:58 | 841 | 3642 | 153 | 142 | 3.8 | 0.12 | 13.43 | 13.11 | 4.75 | 0.05-0.1 |
| | I07AU | 16:42:03 | 250 | 4668 | 171 | 166 | 3.8 | 0.08 | 11.60 | 9.10 | 1.22 | 0.03-0.15 |
| **8-Mar-10** [22:02:07] | I10CA | 01:21:20 | 299 | 2041 | 172 | 246 | 0.85 | 0.01 | 0.81 | 1.10 | 3.22 | 0.7-3 |
| **28-Feb-10** [22:24:50] | I26DE | 23:04:18 | 111 | 534 | 89 | 88 | 0.44 | 0.36 | 2.68 | 2.73 | 2.55 | 0.3-1.5 |
| | I43RU | 23:38:58 | 559 | 1404 | 238 | 235 | 0.44 | 0.68 | 2.49 | 2.59 | 6.61 | 0.3-3 |
| | I31KZ | 00:45:11 | 1139 | 2651 | 280 | 276 | 0.44 | 0.32 | 2.15 | 2.51 | 22.70 | 0.35-3.5 |
| | I46RU | 02:29:27 | 833 | 4320 | 289 | 274 | 0.44 | 0.13 | 2.98 | 3.11 | 2.49 | 0.3-2 |
| **15-Jan-10** [19:17:54] | I32KE | 20:35:00 | 488 | 1342 | 234 | 231 | 1.2 | 0.07 | 2.07 | 2.31 | 4.88 | 0.4-3.5 |
| | I35NA | 20:40:38 | 596 | 1581 | 41 | 35 | 1.2 | 0.48 | 3.08 | 3.85 | 36.82 | 0.2-3.5 |
| | I47ZA | 21:27:37 | 241 | 2267 | 5 | 2 | 1.2 | 0.06 | 1.76 | 1.93 | 2.61 | 0.4-2 |
| **21-Nov-09** [20:53:00] | I35NA | 21:58:54 | 568 | 1249 | 107 | 107 | 18 | 0.48 | 3.47 | 3.98 | 27.22 | 0.2-8 |
| | I49GB | 00:56:30 | 60 | 4312 | 79 | 76 | 18 | 3.17 | 12.02 | 10.24 | 1.70 | 0.05-1 |
| | I17CI | 01:29:58 | 109 | 4892 | 132 | 132 | 18 | 0.01 | 1.35 | 1.37 | 1.33 | 0.6-1.5 |
| **08-Oct-09** [02:57:00] | I39PW | 04:50:53 | 151 | 2024 | 230 | 274 | 33 | 1.12 | 6.23 | 6.40 | 0.48 | 0.08-1.5 |
| | I07AU | 04:56:20 | 925 | 2296 | 318 | 317 | 33 | 1.13 | 3.18 | 3.25 | 2.73 | 0.4-6 |
| | I04AU | 05:59:50 | 954 | 3408 | 8 | 9 | 33 | 0.20 | 3.60 | 2.66 | 9.26 | 0.3-6 |
| | I30JP | 07:37:47 | 457 | 4851 | 209 | 211 | 33 | 0.11 | 1.79 | 1.68 | 2.19 | 0.5-1.5 |
| | I05AU | 07:48:12 | 411 | 5030 | 320 | 317 | 33 | 0.31 | 3.88 | 3.53 | 2.50 | 0.2-2.5 |
| | I22FR | 07:50:18 | 649 | 5362 | 285 | 286 | 33 | 0.14 | 4.06 | 3.79 | 1.66 | 0.1-2 |
| | I45RU | 08:11:17 | 755 | 5500 | 195 | 197 | 33 | 1.18 | 10.46 | 9.87 | 5.85 | 0.05-2 |
| | I44RU | 09:50:22 | 914 | 7254 | 222 | 222 | 33 | 0.62 | 6.58 | 7.55 | 20.65 | 0.1-1.5 |
| | I55US | 10:58:30 | 475 | 8620 | 312 | 315 | 33 | 0.12 | 6.14 | 6.16 | 2.31 | 0.1-1 |
| | I53US | 12:50:45 | 551 | 10503 | 270 | 270 | 33 | 0.43 | 11.78 | 12.60 | 7.00 | 0.05-1 |
| | I18DK | 14:19:35 | 671 | 11816 | 350 | 338 | 33 | 0.48 | 11.50 | 11.54 | 3.24 | 0.08-1 |
| | I56US | 15:04:50 | 645 | 12693 | 293 | 319 | 33 | 0.76 | 13.83 | 12.80 | 33.55 | 0.05-1.5 |
| **04-Sep-09** [02:23:18] | I34MN | 17:06:08 | 484 | 653 | 153 | 86 | 2.3 | 0.10 | 7.50 | 7.45 | 1.23 | 0.08-1 |
| **23-Aug-09** [21:17:19] | I27DE | 22:31:17 | 380 | 1092 | 85 | 85 | 0.75 | 0.13 | 2.44 | 3.15 | 13.33 | 0.35-3 |
| **10-Apr-09** [18:42:45] | I47ZA | 20:22:29 | 87 | 1788 | 179 | 174 | 0.73 | 0.02 | 1.90 | 2.18 | 1.46 | 0.4-1.5 |
| **07-Feb-09** [19:51:32] | I46RU | 20:41:53 | 562 | 994 | 293 | 296 | 3.5 | 0.29 | 1.93 | 2.37 | 56.58 | 0.4-3 |
| | I34MN | 22:15:51 | 342 | 2641 | 305 | 250 | 3.5 | 0.38 | 3.32 | 2.56 | 11.62 | 0.1-2 |
| | I45RU | 23:53:50 | 580 | 4446 | 311 | 296 | 3.5 | 0.06 | 2.38 | 2.07 | 3.27 | 0.4-1.5 |
| | I18DK | 00:31:39 | 813 | 4817 | 32 | 23 | 3.5 | 0.09 | 3.59 | 2.81 | 4.24 | 0.3-2 |
| | I44RU | 00:58:43 | 339 | 5246 | 311 | 304 | 3.5 | 0.05 | 2.11 | 1.52 | 1.35 | 0.5-2 |
| | I53US | 01:46:14 | 397 | 6137 | 3352 | 336 | 3.5 | 0.31 | 6.90 | 7.06 | 3.54 | 0.1-2 |
| **21-Nov-08** [00:26:44] | I56US | 01:13:57 | 159 | 740 | 41 | 42 | 0.41 | 0.28 | 3.73 | 3.66 | 1.40 | 0.1-1.66 |
| | I10CA | 01:17:40 | 420 | 1009 | 294 | 291 | 0.41 | 1.57 | 3.95 | 3.66 | 48.24 | 0.1-2 |
| | I18DK | 03:27:19 | 658 | 3177 | 235 | 225 | 0.41 | 0.47 | 4.56 | 5.46 | 15.83 | 0.15-3.8 |
| **07-Oct-08** [02:45:45] | I32KE | 05:08:16 | 266 | 2532 | 347 | 347 | 1 | 0.03 | 3.31 | 4.18 | 2.54 | 0.2-0.8 |
| | I31KZ | 06:22:36 | 556 | 4024 | 225 | 222 | 1 | 0.13 | 3.39 | 3.47 | 1.36 | 0.2-3 |
| **23-Jul-08** [14:45:25] | I31KZ | 16:03:25 | 569 | 1530 | 145 | 146 | 0.36 | 0.16 | 3.12 | 2.45 | 8.84 | 0.2-5.8 |
| | I46RU | 16:42:09 | 381 | 2130 | 224 | 227 | 0.36 | 0.05 | 2.45 | 3.01 | 2.60 | 0.3-0.85 |
| **08-Jul-08** [15:55:30] | I53US | 18:25:22 | 298 | 2594 | 318 | 319 | 0.21 | 0.01 | 3.27 | 3.20 | 1.35 | 0.3-1.2 |
| **01-Jul-08** [17:40:19] | I53US | 21:22:51 | 151 | 3737 | 130 | 141 | 0.12 | 0.03 | 1.69 | 1.40 | 3.62 | 0.4-3 |
| **27-Jun-08** [02:01:23] | I49GB | 03:04:07 | 449 | 1240 | 335 | 339 | 0.49 | 0.41 | 2.74 | 2.12 | 4.73 | 0.3-1.5 |
| **17-Feb-08** [12:19:16] | I18DK | 12:40:01 | 287 | 334 | 201 | 191 | 0.33 | 0.11 | 3.05 | 3.18 | 2.44 | 0.3-2.5 |
| **09-Jan-08** [03:53:15] | I02AR | 05:03:59 | 564 | 1359 | 180 | 177 | 0.14 | 0.19 | 1.67 | 1.65 | 4.42 | 0.5-5 |
| **12-Oct-07** [09:14:03] | I18DK | 10:42:15 | 217 | 1528 | 359 | 355 | 0.11 | 0.08 | 2.20 | 2.38 | 1.46 | 0.4-2 |
| | I53US | 11:49:39 | 520 | 2815 | 357 | 358 | 0.11 | 0.23 | 4.89 | 4.48 | 11.61 | 0.1-3.5 |
| **22-Sep-07** [17:57:12] | I02AR | 19:09:41 | 341 | 1379 | 288 | 296 | 0.65 | 1.57 | 2.51 | 3.06 | 36.95 | 0.3-5 |
| **18-Apr-07** [12:44:23] | I55US | 13:27:05 | 373 | 754 | 160 | 156 | 0.33 | 1.43 | 6.89 | 6.83 | 69.60 | 0.05-9 |
| **17-Mar-07** [06:48:35] | I17CI | 07:43:53 | 262 | 990 | 87 | 88 | 0.13 | 0.11 | 1.64 | 1.59 | 4.53 | 0.2-2 |
| **22-Jan-07** [07:24:56] | I31KZ | 07:52:34 | 265 | 651 | 213 | 206 | 0.24 | 2.19 | 3.22 | 3.33 | 11.41 | 0.3-9 |
| | I46RU | 09:28:05 | 514 | 2421 | 260 | 261 | 0.24 | 0.41 | 3.02 | 2.48 | 19.84 | 0.2-2 |
| **09-Dec-06** [06:31:12] | I26DE | 09:09:11 | 251 | 2733 | 153 | 150 | 14 | 0.39 | 5.37 | 4.88 | 2.42 | 0.05-1 |
| | I35NA | 11:15:33 | 1350 | 5129 | 11 | 133 | 14 | 0.16 | 5.49 | 4.89 | 1.80 | 0.2-1.5 |
| **14-Oct-06** [18:10:49] | I53US | 20:22:29 | 661 | 2340 | 236 | 236 | 0.7 | 0.04 | 2.87 | 2.16 | 3.11 | 0.4-1.5 |
| **02-Sep-06** [04:26:15] | I04AU | 06:34:36 | 761 | 2404 | 341 | 342 | 2.8 | 1.67 | 7.14 | 6.61 | 33.96 | 0.05-6 |
| | I07AU | 07:06:08 | 433 | 2760 | 280 | 277 | 2.8 | 0.17 | 1.55 | 1.77 | 2.04 | 0.4-4 |
| | I22FR | 10:20:04 | 256 | 6137 | 268 | 261 | 2.8 | 0.23 | 10.68 | 10.78 | 3.30 | 0.07-0.2 |
| **15-Jul-06** [23:55:45] | I55US | 02:41:20 | 239 | 2660 | 184 | 188 | 0.26 | 0.07 | 2.86 | 2.88 | 3.49 | 0.3-3 |
| **07-Jun-06** | I26DE | 02:09:44 | 504 | 2313 | 9 | 8 | 0.19 | 0.03 | 2.95 | 2.38 | 3.50 | 0.2-2 |
| [00:06:28] | I31KZ | 02:46:15 | 429 | 2806 | 331 | 335 | 0.19 | 0.02 | 4.73 | 4.71 | 2.31 | 0.1-1.5 |
| **28-Jan-06** [03:33:48] | I47ZA | 06:59:36 | 210 | 3637 | 144 | 144 | 1.8 | 0.02 | 1.54 | 1.37 | 1.09 | 0.5-1.5 |
| | I33MG | 07:04:01 | 372 | 3721 | 170 | 171 | 1.8 | 0.09 | 2.16 | 2.24 | 2.07 | 0.5-2 |

Table S2. Summary of bolides infrasound signal measurements of 50 individual bolide events as detected from 88 infrasound stations from Ens et al. (2012), not included in Table S1. These measurements are strictly from Ens et al. (2012) except for two events September 3, 2004 and March 27, 2003 where we have independently repeated the inframeasure analysis.

| **Date/ [Time]** | **Station** | **Arrival time (UT)** | **Duration (s)** | **Range (km)** | **Theo. Az (deg)** | **Obs. Az (deg)** | **Satellite Energy (kT)** | **P2P Amp (Pa)** | **Period @Max Amp (s)** | **Period @Max PSD (s)** | **Bolide Integrated Energy SNR** | **Band-pass (Hz)** |
|---|---|---|---|---|---|---|---|---|---|---|---|---|
| **22-Jul-08** [19:34:00] | IS09BR | 23:32:06 | 765 | 4384 | 261 | 248 | 0.25 | 0.12 | 6.69 | 7.25 | 2.29 | 0.08-0.65 |
| **07-Apr-08** [01:02:28] | IS41PY | 1:35:41 | 248 | 618 | 191 | 192 | 0.08 | 0.17 | 1.69 | 2.38 | 4.99 | 0.31-9.8 |
| | IS09BR | 3:00:34 | 413 | 2087 | 209 | 209 | 0.08 | 0.11 | 4.28 | 3.85 | 2.19 | 0.13-1.8 |
| **10-Mar-08** [17:16:08] | IS56US | 19:38:11 | 62 | 2724 | 75 | 82 | 0.05 | 0.10 | 1.20 | 1.74 | 2.88 | 0.3-3.5 |
| **26-Dec-07** [06:46:20] | IS05AU | 10:19:02 | 432 | 4066 | 140 | 131 | 0.4 | 0.11 | 4.32 | 4.63 | 0.95 | 0.2-0.7 |
| **07-Nov-07** [22:31:27] | IS33MG | 1:20:26 | 219 | 3023 | 36 | 35 | 0.05 | 0.02 | 3.11 | 2.77 | 1.62 | 0.25-1.13 |
| **24-Oct-07** [20:28:29] | IS52GB | 21:52:43 | 198 | 1497 | 38 | 31 | 0.02 | 0.08 | 3.13 | 3.72 | 1.05 | 0.2-1.8 |
| **22-Oct-07** [20:10:30] | IS31KZ | 23:34:32 | 418 | 3711 | 313 | 305 | 0.04 | 0.05 | 2.51 | 2.08 | 1.99 | 0.32-1.6 |
| **28-May-07** [20:46:39] | IS46RU | 22:15:15 | 192 | 1524 | 30 | 7 | 0.05 | 0.07 | 2.72 | 2.38 | 1.08 | 0.28-1.5 |
| | IS31KZ | 23:11:19 | 390 | 2950 | 41 | 30 | 0.05 | 0.03 | 2.16 | 2.37 | 5.24 | 0.21-3 |
| **16-May-07** [16:20:58] | IS39PW | 18:50:16 | 1863 | 2866 | 243 | 242 | 0.16 | 0.07 | 2.18 | 2.99 | 1.93 | 0.3-2.5 |
| **16-May-07** [21:28:54] | IS26DE | 22:14:31 | 296 | 788 | 79 | 83 | 0.02 | 0.17 | 0.87 | 1.08 | 3.70 | 0.18-7.5 |
| **09-Feb-02** [19:50:26] | IS07AU | 20:20:07 | 277 | 384 | 92 | 77 | 0.16 | 1.65 | 1.71 | 3.20 | 6.51 | 0.09-9.9 |
| **17-Jan-07** [09:50:46] | IS35NA | 13:18:06 | 66 | 3768 | 77 | 80 | 1.36 | 0.11 | 3.11 | 3.30 | 0.94 | 0.2-2 |
| | IS33MG | 10:57:45 | 183 | 1212 | 18 | 14 | 1.36 | 0.45 | 3.16 | 3.59 | 2.63 | 0.16-0.9 |
| | IS32KE | 11:20:59 | 593 | 1757 | 118 | 121 | 1.36 | 0.64 | 2.74 | 3.62 | 6.24 | 0.17-8 |
| **06-Dec-06** [07:51:07] | IS07AU | 9:19:05 | 40 | 1370 | 226 | 232 | 0.02 | 0.05 | 2.11 | 3.41 | 3.12 | 0.09-1.3 |
| | IS04AU | 8:45:58 | 261 | 1036 | 54 | 48 | 0.02 | 0.14 | 2.17 | 2.59 | 5.79 | 0.2-8 |
| **1-Dec-06** [06:09:25] | IS47ZA | 7:48:59 | 199 | 1784 | 19 | 18 | 0.21 | 0.12 | 1.96 | 1.39 | 2.60 | 0.4-1.9 |
| | IS35NA | 7:30:50 | 280 | 1540 | 68 | 65 | 0.21 | 0.17 | 1.71 | 2.93 | 2.44 | 0.44-1.9 |
| | IS32KE | 7:34:03 | 513 | 1511 | 206 | 208 | 0.21 | 0.06 | 3.50 | 4.68 | 1.49 | 0.16-2.6 |
| **02-Oct-06** [19:10:27] | IS33MG | 22:35:52 | 958 | 3725 | 10 | 11 | 0.25 | 0.12 | 3.27 | 3.51 | 3.94 | 0.08-1.4 |
| | IS32KE | 21:31:54 | 582 | 2467 | 46 | 45 | 0.25 | 0.07 | 5.39 | 4.12 | 3.30 | 0.11-1.8 |
| **11-Sep-06** [12:41:22] | IS53US | 13:27:13 | 398 | 841 | 243 | 237 | 0.09 | 0.18 | 1.79 | 2.17 | 15.15 | 0.16-4.5 |
| **17-Aug-06** [10:43:34] | IS39PW | 14:25:06 | 485 | 4029 | 121 | 114 | 0.63 | 0.07 | 1.70 | 1.57 | 2.36 | 0.48-2.2 |
| | IS22FR | 11:51:16 | 243 | 1221 | 357 | 353 | 0.63 | 0.26 | 3.22 | 2.73 | 1.70 | 0.3-4 |
| **15-Aug-06** [10:52:24] | IS32KE | 12:16:43 | 708 | 1575 | 64 | 67 | 0.16 | 0.07 | 2.05 | 2.16 | 0.03 | 0.305-3 |
| **09-Aug-06** [04:30:44] | IS33MG | 5:12:52 | 556 | 804 | 128 | 124 | 2.32 | 1.32 | 5.98 | 6.11 | 1.90 | 0.09-0.7 |
| | IS32KE | 7:16:13 | 798 | 3037 | 144 | 147 | 2.32 | 0.20 | 5.12 | 4.58 | 5.63 | 0.04-2.3 |
| | IS26DE | 12:47:43 | 595 | 8929 | 143 | 146 | 2.32 | 0.05 | 10.72 | 8.90 | 0.60 | 0.072-0.45 |
| **16-Jul-06** [23:55:45] | IS55US | 2:41:43 | 116 | 2660 | 187 | 189 | 0.27 | 0.12 | 3.61 | 4.02 | 2.07 | 0.15-1 |
| **21-May-06** | IS50GB | 9:02:04 | 537 | 1283 | 251 | 251 | 0.6 | 0.39 | 3.96 | 3.05 | 2.04 | 0.3-4 |
| [07:51:11] | IS17CI | 10:35:47 | 3218 | 3055 | 228 | 228 | 0.6 | 0.20 | 4.21 | 8.45 | 5.87 | 0.07-2.5 |
| | IS09BR | 10:18:27 | 69 | 2475 | 82 | 78 | 0.6 | 0.03 | 3.15 | 5.39 | 2.73 | 0.16-2.3 |
| **14-Mar-06** [03:21:06] | IS08BO | 5:47:47 | 389 | 2664 | 9 | 11 | 0.3 | 0.09 | 3.74 | 4.93 | 7.74 | 0.18-0.4 |
| **06-Feb-06** [01:57:37] | IS35NA | 5:39:16 | 674 | 3923 | 179 | 177 | 2.93 | 0.09 | 9.29 | 7.95 | 2.73 | 0.05-0.3 |
| **09-Nov-05** [07:33:08] | IS22FR | 9:43:27 | 851 | 2291 | 240 | 237 | 0.66 | 0.09 | 9.81 | 8.07 | 1.85 | 0.1-0.9 |
| | IS07AU | 9:07:59 | 319 | 1689 | 139 | 141 | 0.66 | 0.23 | 2.53 | 2.51 | 1.41 | 0.31-3.2 |
| | IS05AU | 8:43:53 | 402 | 1291 | 354 | 352 | 0.66 | 0.09 | 2.73 | 2.41 | 3.26 | 0.29-1.3 |
| **26-Oct-05** [21:30:47] | IS41PY | 23:44:51 | 728 | 2460 | 237 | 237 | 0.43 | 0.10 | 6.45 | 6.45 | 4.09 | 0.1-1.1 |
| | IS14CL | 21:48:21 | 318 | 334 | 177 | 206 | 0.43 | 0.97 | 4.80 | 6.02 | 7.90 | 0.1-9.3 |
| | IS08BO | 23:53:54 | 447 | 2532 | 207 | 205 | 0.43 | 0.16 | 4.27 | 3.86 | 3.17 | 0.11-1.2 |
| **04-Sep-05** [23:04:36] | IS49GB | 0:38:42 | 297 | 1862 | 277 | 274 | 0.14 | 0.35 | 1.89 | 1.55 | 4.52 | 0.3-4 |
| | IS09BR | 1:23:16 | 448 | 2514 | 144 | 140 | 0.14 | 0.12 | 2.25 | 2.02 | 2.90 | 0.28-3.5 |
| **15-Apr-05** [06:54:59] | IS41PY | 9:02:14 | 594 | 2446 | 213 | 225 | 0.08 | 0.04 | 2.07 | 1.87 | 2.21 | 0.36-2.2 |
| **06-Mar-05** [14:12:23] | IS35NA | 14:57:40 | 358 | 835 | 27 | 25 | 0.09 | 0.14 | 1.08 | 1.28 | 4.53 | 0.29-4.9 |
| | IS32KE | 16:18:16 | 455 | 2148 | 234 | 244 | 0.09 | 0.06 | 3.00 | 2.69 | 2.80 | 0.21-0.7 |
| **01-Jan-05** [03:44:09] | IS31KZ | 7:34:08 | 241 | 4191 | 260 | 254 | 0.94 | 0.12 | 3.38 | 3.76 | 2.13 | 0.2-1.8 |
| **29-Dec-04** [07:11:45] | IS22FR | 9:34:35 | 586 | 2743 | 61 | 58 | 0.15 | 0.15 | 3.58 | 5.29 | 2.51 | 0.21-3.6 |
| **11-Dec-04** [15:36:51] | IS34MN | 16:50:12 | 379 | 1437 | 190 | 182 | 0.09 | 0.20 | 3.06 | 3.10 | 4.74 | 0.218-5 |
| **05-Dec-04** [17:14:11] | IS07AU | 19:18:22 | 684 | 2187 | 127 | 125 | 0.07 | 0.16 | 1.92 | 2.07 | 2.99 | 0.22-2.6 |
| | IS05AU | 18:24:01 | 270 | 1382 | 23 | 35 | 0.07 | 0.08 | 2.77 | 2.82 | 1.70 | 0.31-4.5 |
| **07-Oct-04** [13:14:43] | IS55US | 20:03:44 | 471 | 7189 | 263 | 270 | 18.42 | 0.25 | 11.43 | 14.12 | 4.17 | 0.05-0.6 |
| | IS52GB | 15:15:33 | 857 | 2218 | 181 | 177 | 18.42 | 0.28 | 8.96 | 9.81 | 1.26 | 0.089-2 |
| | IS26DE | 23:07:07 | 4230 | 10221 | 131 | 127 | 18.42 | 0.13 | 15.90 | 17.86 | 1.51 | 0.03-0.2 |
| | IS10CA | 6:20:13 | 2374 | 17256 | 27 | 24 | 18.42 | 0.09 | 15.71 | 10.08 | 0.02 | 0.04-0.2 |
| **03-Sep-04** [12:07:24] | IS27DE | 13:18:45 | 1150 | 1088 | 85 | 91 | 13 | 0.82 | 45.72 | 40.96 | 19.53 | 0.015-0.08 |
| | IS55US | 15:54:02 | 1119 | 3716 | 201 | 213 | 13 | 0.91 | 20.66 | 21.85 | 43.79 | 0.04-3 |
| | IS35NA | 17:18:04 | 1212 | 5394 | 180 | 173 | 13 | 0.21 | 13.14 | 14.50 | 1.96 | 0.02-3 |
| **11-Jun-04** [10:51:17] | IS08BO | 12:17:42 | 311 | 1576 | 251 | 242 | 0.06 | 0.05 | 2.93 | 1.38 | 1.80 | 0.298-3 |
| **03-Jun-04** [09:40:12] | IS57US | 11:13:47 | 307 | 1671 | 345 | 342 | 0.14 | 0.07 | 4.09 | 3.15 | 3.75 | 0.14-2.9 |
| | IS56US | 9:59:13 | 161 | 378 | 267 | 264 | 0.14 | 0.05 | 2.97 | 2.73 | 3.53 | 0.2-2 |
| | IS53US | 12:04:01 | 222 | 2417 | 130 | 143 | 0.14 | 0.03 | 1.13 | 1.97 | 3.16 | 0.45-2.7 |
| **22-Apr-04** [21:19:55] | IS33MG | 22:18:00 | 59 | 930 | 65 | 66 | 0.37 | 0.10 | 2.95 | 3.41 | 2.95 | 0.3-3.14 |
| **17-Aug-03** [13:16:07] | IS55US | 19:33:41 | 239 | 6664 | 226 | 237 | 1.62 | 0.04 | 3.93 | 5.69 | 1.64 | 0.18-1.5 |
| | IS33MG | 15:35:40 | 206 | 2565 | 207 | 205 | 1.62 | 0.16 | 3.65 | 4.65 | 1.00 | 0.16-1.1 |
| **27-Mar-03** [05:50:26] | IS10CA | 6:56:00 | 450 | 1143 | 136 | 151 | 0.15 | 0.14 | 2.00 | 2.21 | 7.95 | 0.3-3 |
| **10-Nov-02** [22:13:54] | IS59US | 23:18:19 | 362 | 1142 | 252 | 250 | 1.17 | 4.13 | 5.12 | 6.61 | 123.79 | 0.015866-9.9 |
| **09-Oct-02** [12:00:35] | IS59US | 14:49:31 | 635 | 3427 | 223 | 212 | 0.19 | 0.10 | 5.18 | 4.68 | 1.73 | 0.08-4 |
| **14-Aug-02** [07:48:32] | SSG0 | 12:38:59 | 504 | 5056 | 198 | 207 | 0.27 | 0.06 | 3.78 | 4.76 | 1.70 | 0.1-2.2 |
| | nSts | 12:34:10 | 381 | 4943 | 195 | 199 | 0.27 | 0.06 | 2.61 | 2.45 | 1.98 | 0.15-1.3 |
| | IS59US | 11:51:53 | 574 | 4387 | 130 | 130 | 0.27 | 0.13 | 3.29 | 4.10 | 3.87 | 0.08-1.5 |
| **25-Jul-02** [15:57:32] | IS33MG | 16:56:29 | 236 | 1078 | 183 | 178 | 0.69 | 0.25 | 4.08 | 4.76 | 3.56 | 0.17-3 |
| **13-Jun-02** [15:29:38] | IS07AU | 17:45:05 | 293 | 2417 | 252 | 250 | 0.56 | 0.05 | 3.89 | 3.50 | 2.59 | 0.25-2.3 |
| **06-Jun-02** [04:28:30] | IS26DE | 5:59:48 | 2081 | 1757 | 157 | 160 | 7.58 | 1.66 | 13.33 | 13.16 | 14.29 | 0.025-6.2 |
| **09-Mar-02** [01:20:24] | PSDI | 6:20:20 | 372 | 5431 | 234 | 250 | 0.61 | 0.47 | 5.00 | 4.93 | 2.83 | 0.08-2.2 |
| | nSts | 5:31:42 | 428 | 4583 | 232 | 241 | 0.61 | 0.22 | 4.48 | 4.43 | 1.82 | 0.15-4 |
| | IS59US | 2:58:00 | 489 | 1689 | 147 | 146 | 0.61 | 0.36 | 8.51 | 10.24 | 3.92 | 0.052-5 |
| **23-Jul-01** [22:19:00] | EC1 | 22:37:47 | 267 | 323 | 354 | 11 | 2.87 | 1.62 | 2.82 | 2.77 | 8.61 | 0.1-39 |
| **23-Apr-01** [06:12:00] | SSG0 | 8:09:42 | 442 | 2039 | 248 | 252 | 8.97 | 1.10 | 4.11 | 5.12 | 27.72 | 0.05-7.5 |
| | NSV0 | 7:53:05 | 454 | 1752 | 235 | 237 | 8.97 | 6.23 | 4.67 | 5.18 | 20.68 | 0.05-5 |
| | nSts | 7:55:25 | 863 | 1835 | 245 | 241 | 8.97 | 1.19 | 3.77 | 8.62 | 22.49 | 0.03-6.5 |
| | IS59US | 8:21:55 | 779 | 2526 | 62 | 61 | 8.97 | 0.55 | 4.96 | 4.22 | 7.56 | 0.02-9.3 |
| | IS57US | 7:49:04 | 400 | 1670 | 254 | 255 | 8.97 | 19.07 | 4.27 | 4.14 | 25.45 | 0.03-9.3 |
| | IS26DE | 16:22:12 | 781 | 10810 | 331 | 324 | 8.97 | 0.09 | 7.50 | 6.83 | 1.98 | 0.07-0.8 |
| | DSLI | 8:40:06 | 934 | 2627 | 259 | 261 | 8.97 | 0.26 | 4.01 | 4.46 | 2.59 | 0.1-2 |
| **25-Aug-00** [01:12:25] | DSLI | 3:25:16 | 809 | 2382 | 179 | 185 | 3.15 | 0.07 | 6.77 | 5.61 | 9.74 | 0.04-2.5 |
| **19-Jul-00** [17:40:25] | WSRA | 21:36:46 | 208 | 4231 | 266 | 262 | 0.37 | 0.38 | 3.51 | 3.10 | 0.16 | 0.18-1.5 |

S1. Blast radius plot for the Park Forest meteorite dropping fireball of March 23, 2004.

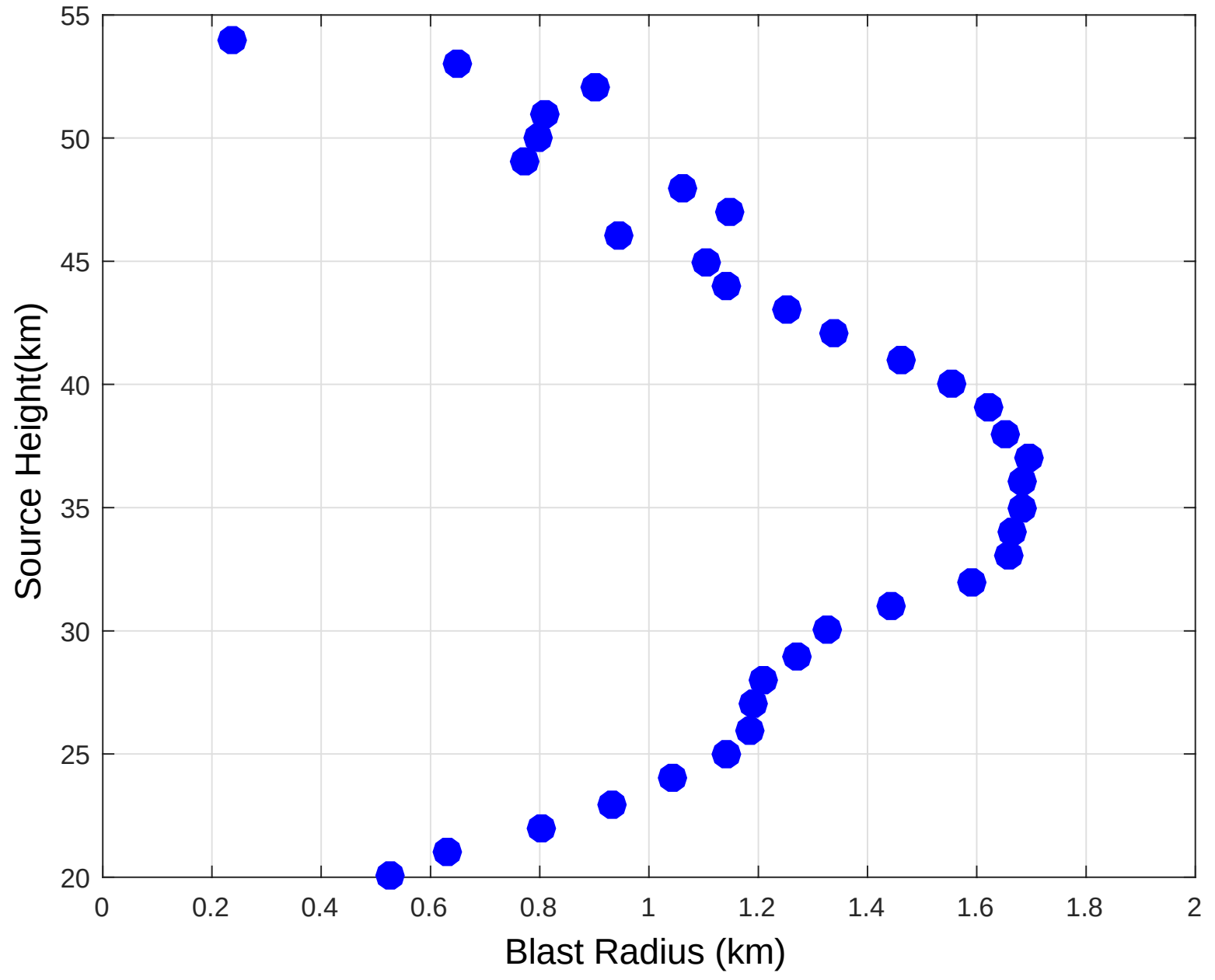